\documentclass[useAMS,usenatbib]{mn2e}

\usepackage{color}
\usepackage{graphicx}
\usepackage{times}
\usepackage{natbib}
\usepackage{amsmath,amssymb,amsfonts,txfonts}
\usepackage[latin1]{inputenc}

\newif\ifAMStwofonts
\AMStwofontstrue

\newcommand{\mincir}{\raise
  -2.truept\hbox{\rlap{\hbox{$\sim$}}\raise5.truept \hbox{$<$}\ }}
\newcommand{\magcir}{\raise
  -2.truept\hbox{\rlap{\hbox{$\sim$}}\raise5.truept \hbox{$>$}\ }}
\newcommand{\siml}{\raise
  -2.truept\hbox{\rlap{\hbox{$\sim$}}\raise5.truept \hbox{$<$}\ }}
\newcommand{\simg}{\raise
  -2.truept\hbox{\rlap{\hbox{$\sim$}}\raise5.truept \hbox{$>$}\ }}

\newcommand{\msun}{\rm{M}_{\odot}}
\newcommand{\rfive}{r_{500,c}}
\newcommand{\rtwo}{r_{200,c}}
\newcommand{\mtwo}{M_{200,c}}
\newcommand{\mfive}{M_{500,c}}
\newcommand{\hab}{H_{AB}}
\newcommand{\ks}{K_s}
\newcommand{\ha}{\rm{H}\alpha}

\newcounter{jwcomment}
\newcommand{\eu}{\emph{Euclid}}

 \newcommand{\Planck}{Planck}

\newcommand{\cpl}{CPL}

\newcommand{\nong}{non-Gaussian}

\newcommand{\de}{DE}

\begin{document}

\title{Next Generation Cosmology: Constraints from the \emph{Euclid} 
Galaxy Cluster Survey}

\author[B.\,Sartoris et al.]{
B.\,Sartoris$^{1,2}$, 
A.\,Biviano$^{2}$, 
C.\,Fedeli$^{3,4}$,
J. G.\,Bartlett$^{5}$,
S.\,Borgani$^{1,2,6}$,
M.\,Costanzi$^{7}$,\\~\\
\LARGE{\rm 
C.\,Giocoli$^{3,4,8,9}$,
L.\,Moscardini$^{8,3,4}$, 
J.\,Weller$^{7,10,11}$,
B.\, Ascaso$^{12}$,
S.\,Bardelli$^{3}$,}\\~\\
\LARGE{\rm 
S.\,Maurogordato$^{13}$,
and P. T. P.\,Viana$^{14,15}$ 
}\\~\\
$^1$ Dipartimento di Fisica, Sezione di Astronomia, Universit\`a di
Trieste, Via Tiepolo 11, I-34143 Trieste, Italy\\ 
$^2$ INAF/Osservatorio Astronomico di Trieste, Via Tiepolo 11, I-34143
Trieste, Italy\\ 
$^3$ INAF/Osservatorio Astronomico di Bologna, Via Ranzani 1, I-40127
Bologna, Italy\\
$^4$ INFN, Sezione di Bologna, Viale Berti Pichat 6/2, I-40127
Bologna, Italy\\
$^5$ APC, AstroParticule et Cosmologie, Universit\'e Paris Diderot, CNRS/IN2P3,
CEA/lrfu,\\ Observatoire de Paris, Sorbonne Paris Cit\'e, 10, rue Alice Domon et
L\'eonie Duquet, 75205 Paris Cedex 13, France\\
$^6$ INFN, Sezione di Trieste, Via Valerio 2, I-34127 Trieste, Italy\\
$^7$ Universit\"ats-Sternwarte M\"unchen, Fakult\"at f\"ur Physik,
           Ludwig-Maximilians Universit\"at M\"unchen, Scheinerstr. 1, D-81679
           Mu\"nchen, Germany\\
$^8$ Dipartimento di Fisica e Astronomia, Alma Mater Studiorum
Universit\`{a} di Bologna, viale Berti Pichat, 6/2, 40127 Bologna,
Italy\\
$^9$ Aix Marseille Universit\'e, CNRS, LAM (Laboratoire d'Astrophysique de
Marseille) UMR 7326, 13388, Marseille, France\\
$^{10}$ Excellence Cluster Universe, Boltzmannstr. 2, D-85748 Garching,
Germany\\
$^{11}$ Max Planck Institute for Extraterrestrial Physics, Giessenbachstr. 1,
D-85748 Garching, Germany\\
$^{12}$ GEPI, Observatoire de Paris, CNRS, Universit\'e Paris Diderot, 61, Avenue de l'Observatoire 75014, Paris  France\\
$^{13}$ Laboratoire J.-L. LAGRANGE, CNRS/UMR 7293, Observatoire de la C\^ote
d'Azur, Universit\'e de Nice Sophia-Antipolis,  06304 Nice Cedex 4, France\\
$^{14}$ Instituto de Astrof\'{i}sica e Ci\^{e}ncias do Espa\c{c}o, Universidade
do Porto, CAUP, Rua das Estrelas, 4150-762 Porto, Portugal\\
$^{15}$ Departamento de F\'{i}sica e Astronomia, Faculdade de Ci\^{e}ncias,
Universidade do Porto, Rua do Campo Alegre 687, 4169-007 Porto, Portugal\\
}

\maketitle

\begin{abstract}
   We study the characteristics of the galaxy cluster samples expected
  from the European Space Agency's \emph{Euclid} satellite and
  forecast constraints on parameters describing a variety of
  cosmological models. The method used in this paper, based on the
  Fisher Matrix approach, is the same one used to provide the
  constraints presented in the \emph{Euclid} Red
  Book \citep{laureijs11}.  We describe the analytical approach to
  compute the selection function of the photometric and spectroscopic
  cluster surveys.  Based on the photometric selection function, we
  forecast the constraints on a number of cosmological parameter sets
  corresponding to different extensions of the standard $\Lambda$CDM
  model, including a redshift-dependent Equation of State for Dark
  Energy, primordial non-Gaussianity, modified gravity and
  non-vanishing neutrino masses. Our results show that \eu\ clusters
  will be extremely powerful in constraining the amplitude of the
  matter power spectrum $\sigma_8$ and the mass density parameter
  $\Omega_{\mathrm{m}}$.  The dynamical evolution of dark energy will be
  constrained to $\Delta w_0 = 0.03$ and $\Delta w_a = 0.2$ with free
  curvature $\Omega_k$, resulting in a $(w_0,w_a)$ Figure of Merit
  (FoM) of 291. Including the \emph{Planck} CMB covariance matrix, thereby
  information on the geometry of the universe, improves the
  constraints to $\Delta w_0 = 0.02$, $\Delta w_a = 0.07$ and a
  FoM$=802$.  The amplitude of primordial non-Gaussianity,
  parametrised by $f_\mathrm{NL}$, will be constrained to $\Delta
  f_\mathrm{NL} \simeq 6.6$ for the local shape scenario, from \eu\
  clusters alone. Using only \eu\ clusters, the
  growth factor parameter $\gamma$, which signals deviations from
  General Relativity, will be constrained to $\Delta \gamma = 0.02$,
  and the neutrino density parameter to
  $\Delta \Omega_{\nu} = 0.0013$ (or $\Delta \sum
  m_{\nu} = 0.01$). We emphasise that
  knowledge of the observable--mass scaling relation 
  will be crucial to constrain cosmological parameters
  from a cluster catalogue.  The \eu\ mission will have a clear
  advantage in this respect, thanks to its imaging and spectroscopic
  capabilities that will enable internal mass calibration from weak
  lensing and the dynamics of cluster galaxies. This information will
  be further complemented by wide-area multi-wavelength external
  cluster surveys that will already be available when \eu\ flies.
\end{abstract}

\section{Introduction}
\label{s:intro}
According to the hierarchical scenario for the formation of cosmic
structures, galaxy clusters are the latest objects to have formed from
the collapse of high density fluctuations filtered on a
typical scale of $\sim 10$ comoving Mpc
\citep[e.g.][]{kravtsov_borgani12}.  Since galaxy clusters provide
information on the growth history of structures and on the underlying
cosmological model in many ways \citep[see, e.g.,][]{allen11},
they have played an important role in delineating the current standard
$\Lambda$CDM cosmological model. As a matter of fact, the number
counts and spatial distribution of these objects have a strong
dependence on a number of cosmological parameters, especially the
amplitude of the mass power spectrum and the matter content of the
Universe. The evolution with redshift of the cluster number density
and correlation function can be employed to break the degeneracy
between these two parameters, and thus can provide constraints on the
cold Dark Matter (DM henceforth) and Dark Energy (DE) density
parameters \citep[e.g.,][]{wang98,haiman01,weller02,battye03,allen11,sartoris12}. Furthermore, a
number
of studies \citep[e.g.,][]{carbone12, costanzi13a, costanzi14} have also shown
that clusters can be used to constrain neutrino properties, because
massive neutrinos would directly influence the growth of cosmic
structure, by suppressing the matter power spectrum on small
scales. More generally, since the evolution of the cluster population
traces the growth rate of density perturbations, large surveys of
clusters extending over a wide redshift interval have the potential of
providing stringent constraints on any cosmological model whose
deviation from $\Lambda$CDM leaves its imprint on this growth.

Over the past decade, surveys of galaxy clusters for cosmological use
have been constructed and analysed, based on observations at different
wavelengths: X-ray
\citep[e.g.][]{borgani01,vikhlinin09c,CL12.1,RA13.1}; sub-mm, through
the \citet{SU72.1} distortion (SZ henceforth,
\citealt{ST08.1,BE13.1,PL13.1,BU12.1}), and optical \citep{RO10.1}
bands. Further improvements can be obtained from the spatial clustering of
galaxy clusters \citep{schuecker03,hutsi10,mana13}. The resulting cosmological
constraints turn out to be
complementary to those of other cosmological probes such as type Ia
supernovae \citep[e.g.,][]{betoule14}, Cosmic Microwave Background
(CMB) radiation
\citep[e.g.,][]{wmap9,planck13_16}, the Baryon Acoustic Oscillations
\citep[BAOs; e.g.,][]{anderson14}, and cosmic shear
\citep[e.g.][]{kitching14}.
These cluster catalogues are however
characterised either by a large number of objects that cover a
relatively small redshift range, or rather small samples that span a
wide redshift range. Ideally, in order to exploit the redshift
leverage with good statistics, one should have access to a survey that
can provide a high number of well characterised clusters over a wide
redshift range.

One future mission that will achieve this goal will be the European
Space Agency (ESA) Cosmic Vision mission \eu\footnote{http://www.euclid-ec.org}  \citep[][]{laureijs11}. Planned for launch in
the year $2020$, \eu\ will study the evolution of the cosmic web up to
redshift $z\sim 2$. Although the experiment is optimised for the
measurement of cosmological Weak Lensing (WL, or cosmic shear) and the
galaxy clustering, \eu\ will also provide data usable for other important
complementary cosmological probes, such as galaxy clusters. Cluster
detection will be possible in three different ways: $i)$ using
photometric data; $ii)$ using spectroscopic data; and $iii)$ through
gravitational (mostly weak) lensing, which may be combined for more efficiency.
In this paper, we will perform
our analyses by using the photometric cluster survey (see
Section \ref{s:clsel}), where the cluster detection method is not
dissimilar from that used to detect low-redshift SDSS
clusters \citep{koester07}. However, thanks to the use of Near
Infrared (NIR) bands, \eu\ will be capable of detecting clusters at
much higher redshifts ($z \sim 2$) over a similarly large area. The
sky coverage of \eu\ will reach $15,000 \deg^2$, almost the entire
extragalactic celestial sphere. The characteristics of the \eu\
spectroscopic survey and its possible use for the calibration of the
mass-observable relation will be discussed in Appendices
A and B, respectively.

One fundamental step for the cosmological exploitation of galaxy
clusters is the definition of the relation between the mass of the
host DM halo and a suitable observable quantity
\citep[e.g.,][]{andreon12,giodini13}. Many efforts have been devoted to
the calibration of the observable-mass scaling relations at different
wave bands
\citep[e.g.][]{arnaud10,planck11_sr,reichert11,rozo11,rykoff12,ettori13,rozo14,
mantz14b}
and in the definition of mass proxies which are at the same time
precise (i.e. characterised by a small scatter in the scaling against
cluster mass) and robust (i.e. relatively insensitive to the details
of cluster astrophysics) \citep[e.g.][]{kravtsov06}.  In the case of
\eu, an internal mass calibration will be performed through the
exploitation of spectroscopic and WL data of the wide \eu\ survey (see
Appendix \ref{app:mass}), and of the deep \eu\ survey of $40 \deg^2$,
2 magnitudes deeper than the wide survey. The deep survey will be
particularly useful in adding constraints on the evolution of the
observable-mass scaling relation at $z>1$.

These \eu\ internal data will provide a precise calibration
of the relation between cluster richness, which characterises
photometrically-identified clusters, and their actual mass.
Furthermore, it will be possible to cross-correlate \eu\ data with
data from other cluster surveys - such as
\emph{eRosita}
\citep{ME12.1}, \emph{XCS} \citep{mehrtens12},
the South Pole Telescope (SPT, \citealt{CA11.1}), and
the Atacama Cosmology Telescope (ACT, \citealt{MA11.1}) - to
further improve the mass calibration of \eu\ clusters.

The aim of this paper is to forecast the strength and the peculiarity
of the \eu\ cluster sample in constraining the parameters describing
different classes of cosmological models that deviate from the
concordance $\Lambda$CDM paradigm. We first consider the case of a
dynamical evolution of the DE component, using the two-parameter
functional form originally proposed by \citet{chevallier01} and
\citet{linder03}. The same parametrisation has been used in the Dark
Energy Task Force reports \citep[DETF;][]{albrecht06,albrecht09} to
estimate the constraining power of different cosmological
experiments. Second, we allow for the primordial mass density
perturbations to have a non-Gaussian distribution. Third, we explore
the effect of deviations from General Relativity (GR) on the linear growth
of density perturbations. Finally, we consider the case of including
massive standard neutrinos.

The structure of this paper is the following. In Section
\ref{s:clsel}, we describe the approach used to estimate the \eu\
cluster selection function of the photometric survey. In
Section \ref{s:meth}, we describe the Fisher Matrix 
approach used to derive constraints from the \eu\ cluster survey on
cosmological parameters. In Section \ref{s:mod}, we briefly describe
the characteristics of the different cosmological models we
consider. In Section \ref{s:res}, we show our results on the number of
clusters that the wide \eu\ survey is expected to detect as a function of
redshift and
the constraints that will be obtained on the cosmological parameters
using the cluster number density and power spectrum.  Finally, we
provide our 
discussion and conclusions in Section
\ref{s:concl}. We present
the analytical derivation of the spectroscopic selection
function in Appendix \ref{app:spec} and the calibration of the cluster
observable-mass relation in Appendix \ref{app:mass}.

\section{Galaxy cluster selection in the \eu\ photometric survey}
\label{s:clsel}

In this Section, we adopt the cosmological parameter values of the
concordance $\Lambda$CDM model from \citet{planck13_16}, $H_0=67\,{\rm
km\,s}^{-1} {\rm Mpc}^{-1}$ for the Hubble constant,
$\Omega_{\mathrm{m}}=0.32$ for the present-day matter density
parameter, and $\Omega_{k}=0$ for the curvature parameter.

To determine the selection function of galaxy clusters in the \eu\
photometric survey, we adopt a phenomenological approach. We start by
adopting an average universal luminosity function (LF hereafter) for
cluster galaxies. \citet{LMS03} evaluated the $\ks$-band LFs of
cluster galaxies out to a radius $\rfive$ for several nearby
clusters. The radius $r_\Delta$ is defined as the radius of the sphere
that encloses an average mass density $\Delta$ times the critical
density of the Universe at the cluster redshift. These cluster LFs
were parametrised using Schechter functions \citep{Schechter76}.  We
adopt the averages of the normalisations and characteristic
luminosities listed in Table $1$ of \citet{LMS03} for the 27 nearby
clusters included in that analysis, corresponding to
$\phi^{\star}=6.4$ Mpc$^{-3}$ and $M^{\star}=-24.85$. Also, following
\citet{LMS03}, we use a faint-end slope $\alpha=-1.1$,
as confirmed in the r-band deep spectroscopic analysis of two nearby
clusters by \citet{rines08}.

\begin{figure}
\includegraphics[width=0.47\textwidth]{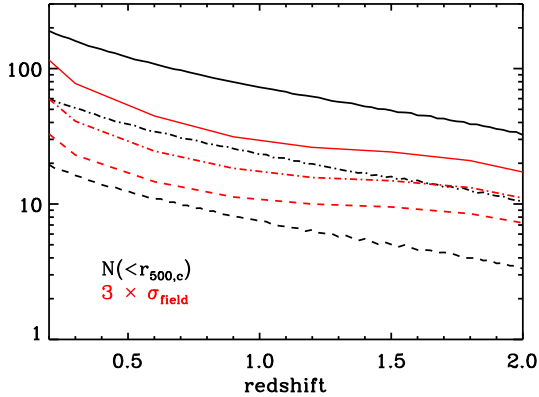}
\caption{Number $N_{500,c}$ of cluster galaxies
within $r_{500,c}$ (black curves), and $3 \sigma_{\rm{field}}$ where
$\sigma_{field}$ is the \emph{rms} of the field counts within the same
radius, and within the adopted $3 \Delta z_\mathrm{p}$ cut (red
curves). These counts are shown down to the limiting magnitude of
the \eu\ survey, $H_{AB} = 24$, as a function of redshift for clusters
of different masses, $\log (M_{200,c}/M_{\odot})=14.5, 14.0,
13.5$ (solid, dot-dashed, dashed line, respectively), where
masses are defined with a mean overdensity of $200$ times the critical
density of the universe at the cluster redshifts.}
\label{f:N500}
\end{figure}

Concerning the behaviour of the cluster LF at $z > 0$, there is no
conclusive observational evidence on the evolution of the LF faint-end
slope parameter $\alpha$ \citep{Mancone+12,SM13}.  Therefore, we
assume it to be redshift-invariant. Also, the observed constancy of
the richness vs. mass relation for clusters up to $z\simeq 0.9$
\citep{Lin+06,Poggianti+10,AC14} suggests that there is no redshift evolution of
$\phi^{\star}$, apart from the cosmological evolution of the critical
density, which scales as $H^2(z)$. 

We assume the $M^{\star}$ parameter
to change with $z$ according to passive evolution models of
stellar populations
\citep{KA97}.  This assumption is justified because emission in the $\ks$ band
is not strongly influenced by young stellar generations, and it is
supported by observations \citep[][and references
therein]{Mancone+12}, at least for clusters more massive than $\sim
10^{14} \, \msun$. For clusters of lower mass, some high-$z$ surveys
have found evidence for deviation from passive evolution of
$M^{\star}$ \citep{Mancone+10,Tran+10,Brodwin+13}. However, the
current observational evidence does not allow us to precisely
parametrise $M^{\star}$ evolution to $z>1$ and low cluster masses, and
we prefer to keep our conservative assumption of passive evolution
over the full cluster mass range.

We apply the
early-type $k$-correction of \citet{Mannucci+01} to the $M^{\star}$ magnitudes.
This correction should be the most appropriate for galaxies in clusters, which
are mostly early-type even at relatively high redshifts
\citep{Postman+05,Smith+05}. We finally convert the $\ks$ magnitudes into the
\eu\ band $\hab$ using the mean rest-frame colour for cluster galaxies, 
$H-\ks=0.26$
\citep[we average the values provided by][]{Boselli+97,dPESD98,Ramella+04}, and
adopting the transformation to the AB-system
$\hab=H+1.37$ \citep{Ciliegi+05}. We thus obtain the cluster LFs
in the $\hab$ band at different redshifts.

By integrating these LFs down to the apparent magnitude limit of
the wide \eu\ photometric survey \citep[$\hab=24$, see][]{laureijs11}, we
then evaluate $n_{500,c}$, namely the redshift-dependent number density
of cluster galaxies within $\rfive$.  The number of cluster galaxies
contained within a sphere of radius $\rfive$ (i.e. the cluster
richness) is then $N_{500,c} = 4 \pi\,n_{500,c}\rfive^3 /3 = 8/3 \pi\,
n_{500,c} G \mfive /[500 H^2(z)]$, where the last equivalence follows
from the relation between $\rfive$ and $\mfive$, the mass within a
mean overdensity of 500 times the critical density of the universe at
the cluster redshifts.  Note that the dependence of $N_{500,c}$ on
$H^{-2}(z)$ is only apparent, since $\phi^{\star}$, and hence
$n_{500,c}$, scales as $H^2(z)$. The $z$-dependence only comes in as a
result of the fixed magnitude limit of the survey and the passive
evolution of galaxies.
In Fig. \ref{f:N500} we show $N_{500,c}(z)$ for clusters
of three different masses: $\log \mtwo=13.5, 14.0,$ and 14.5 (black
curves). To convert from $\mfive$ to $\mtwo$ we adopt a NFW
profile \citep{navarro97} with a mass- and redshift-dependent
concentration given by the relation of
\citet[][2$^{nd}$ relation from top in their Table 5]{DeBoni+13}.

We then
estimate the contamination by field galaxies in the cluster area.  We
take the estimate of the number density of field galaxies down to
$\hab=24$ from the H-band counts of \citet[][see their Table
3]{Metcalfe+06}, $n_\mathrm{field} \simeq 33$ arcmin$^{-2}$, 
an estimate that is in
agreement with the \eu\ survey requirements \citep{laureijs11}.
Multiplying this density by the area subtended by a galaxy cluster at
any given redshift we obtain the number of field galaxies that
contaminate the cluster field-of-view, $N_\mathrm{field} =
n_\mathrm{field}\,\pi\rfive^2$, where $\rfive$ is in arcmin. 

\begin{figure}
\begin{center}
\begin{minipage}{0.5\textwidth}
\resizebox{\hsize}{!}{\includegraphics{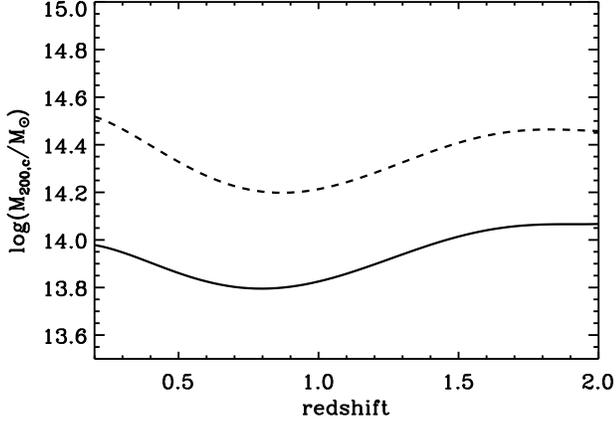}}
\end{minipage}
\end{center}
\caption{Galaxy cluster mass selection function for the \eu\
  photometric survey. Solid and dashed lines are for detection
  thresholds $N_{500,c}/\sigma_{\rm{field}}=3$ and $5$,
  respectively.}
\label{f:selfuncphot}
\end{figure}

The number of field contaminants can be greatly reduced by using
photometric redshifts, $z_\mathrm{p}$. These will be obtained to the
required accuracy of $\Delta z_\mathrm{p} \equiv 0.05 
(1+z_\mathrm{c})$, by combining the \eu\ photometric survey with
auxiliary ground-based data \citep{laureijs11}.  One can safely
consider as non-cluster members all those galaxies that are more than
$3  \Delta z_\mathrm{p}$ away from the mean cluster redshift
$z_\mathrm{c}$. The mean cluster redshift will be evaluated by
averaging the photometric redshifts of galaxies in the cluster region,
and additionally including the (few) spectroscopic galaxy redshifts
provided by the
\eu\ spectroscopic survey (see Appendix~\ref{app:spec}).

In order to determine the fraction of field galaxies,
$f(z_\mathrm{c})$, with photometric redshift $z_\mathrm{p}$ in the
range $\pm 3 \times 0.05 (1+z_\mathrm{c})$ at any given
$z_\mathrm{c}$, we need to estimate the photometric redshift distribution of an
$\hab=24$ limited field survey. To this aim we consider the
photometric redshift distribution of galaxies with $\hab \leq 24$ in
the catalogue of \citet{Yang+14}.  We find $f(z_\mathrm{c})= 0.07, 0.23,
0.34$, and $0.33$ at $z_{\rm{c}}=0.2, 0.8, 1.4$, and $2.0$,
respectively.

Finally, we evaluate the \emph{rms}, $\sigma_\mathrm{field}$, of the
field galaxy counts $f(z_\mathrm{c}) N_\mathrm{field}$, by taking into
account both Poisson noise and cosmic variance.  For the latter we use
the IDL code \texttt{quickcv} of John
Moustakas\footnote{
https://code.google.com/p/idl-moustakas/source/browse/trunk/impro/cosmo/
\newline quickcv.pro?r=617}
for cosmic variance calculation. In Fig. \ref{f:N500}
we show $3 \sigma_{\rm{field}}$ as a function of redshift,
in clusters of $\log \mtwo = 13.5, 14.0,$ and 14.5.

The ratio between the cluster galaxy
number counts and the field
\emph{rms}, $N_{500,c}/\sigma_{\rm{field}}$, gives the significance of
the detection for a given cluster. The cluster selection function is
the limiting cluster mass as a function of redshift for a given
detection threshold. This is shown in Fig. \ref{f:selfuncphot} for two
thresholds, $N_{500,c}/\sigma_{\rm{field}}=3,$ and 5. This selection
function is only mildly dependent on redshift.  The limiting cluster
mass for the lowest selection threshold
($N_{500,c}/\sigma_{\rm{field}}=3$) is $\mtwo \sim 8 \times
10^{13} \, \msun$, lower than the typical mass of richness class $0$
clusters in the \citet{ACO89} catalogue \citep{Popesso+12}.  It is also
similar to the limiting mass of the selection function of SDSS
clusters identified by the maxBCG algorithm \citep[see Fig. 3
in][]{RO10.1}, and to the typical mass of the clusters identified
by \citet{Brodwin+07} up to $z \sim 1.5$
using $z_p$ in an IR-selected galaxy catalogue.
Preliminary tests based on running cluster finders
on \eu\
mocks\footnote{http://wiki.cosmos.esa.int/euclid/index.php/EC\_SGS\_OU\_LE3.
Access restricted to members of the \eu\ Consortium.}, show that the
mass limit $\mtwo \sim 8 \times 10^{13} \, \msun$ roughly corresponds
to $\sim 80$\% completeness at all redshifts $z \leq 2$.

The shape of the selection functions shown in Fig.~\ref{f:selfuncphot}
is somewhat counter-intuitive because it is higher at $z \sim 0.2$
than at $z \sim 0.7$. Naively one would expect that clusters of lower
mass would be easier to detect at lower redshifts. We find that this
shape is related to the relative importance of cosmic variance and
Poisson noise in the contaminating field counts. Cosmic variance
drives the shape of the selection function at $z<0.5$ and Poisson
noise at higher redshifts. If we select clusters at a higher
overdensity (e.g. $\Delta_c=2500$ rather than $\Delta_c=500$), the
relative importance of cosmic variance and Poisson noise changes in a
way to flatten the selection function at $z<0.5$. In reality,
observers do not select clusters at given $\Delta_c$, so our estimate
of the selection function must be considered only as an approximation.
At the end of Section~\ref{s:res} we comment on the effect of taking a
flat selection function out to $z=2$.

\section{Fisher Matrix analysis}
\label{s:meth}
Before presenting our forecasts for the cosmological constraints
we now briefly describe the Fisher Matrix (FM hereafter)
formalism that we use to derive these constraints.

The FM formalism is a Gaussian approximation of the likelihood around the
maximum to second order and it is an efficient way to study the accuracy of the
estimation of a vector of parameters $\boldsymbol p$ by using
independent data sets.  The FM is defined as
\begin{equation}
 F_{\alpha \beta} \equiv  - \left \langle \frac {\partial^2 \ln  {\cal L}}
{\partial p_{\alpha} \partial p_{\beta}} \right \rangle \, ,
\end{equation}
where ${\cal L }$ is the likelihood of the observable quantity
\citep[e.g.][]{dodelson03}. In our FM analysis
we combine three different pieces of information: the galaxy cluster
number density, the cluster power spectrum, and the prior knowledge of
cosmological parameters as derived from the \emph{Planck} CMB
experiment \citep{planck13_16}.  To quantify the constraining power of
a given cosmological probe on a pair of joint parameters $(p_i,p_j)$
we use the Figure of Merit \citep[FoM henceforth;][]{albrecht06}
\begin{equation}
\mathrm{FoM} = \frac{1}{\sqrt{\det\left[\mathrm{Cov}\left( p_i,p_j \right) \right]}}\;,
\label{eq:fom}
\end{equation}
where $Cov(p_i,p_j)$ is the covariance matrix between the two
parameters. With this definition, the FoM is proportional to the
inverse of the area encompassed by the ellipse representing the $68$
per cent confidence level (c.l.) for model exclusion.

As described in detail in \citet{sartoris10}, we follow the approach of
\citet{holder01} and define the FM for the cluster number counts as
\begin{equation}
  F^N_{\alpha \beta}= \sum_{\ell,m} \frac{\partial N_{\ell,m}}{\partial
    p_\alpha}\frac{\partial N_{\ell,m}}{\partial p_\beta}
  \frac{1}{N_{\ell,m}}\,.
\label{eq:fm_nc}
\end{equation}
In the previous equation, the sums over $\ell$ and $m$ run over
redshift and mass intervals, respectively. The quantity $N_{\ell,m}$
is the number of clusters expected in a survey with a sky coverage
$\Omega_\mathrm{sky}$, within the $\ell$-th redshift bin and $m$-th
bin in observed mass $M^\mathrm{ob}$. This can be calculated as \citep{lima05}
\begin{eqnarray}
N_{\ell,m} & = & \frac{\Omega_\mathrm{sky}}{4\pi}
\int_{z_\ell}^{z_{\ell+1}}dz\, 
\frac{dV}{dz} \nonumber \\ 
& & \int_{M^\mathrm{ob}_{\ell,m}}^{M^\mathrm{ob}_{\ell,m+1}}
\frac{dM^\mathrm{ob}}{M^\mathrm{ob}}  \int_0^{+\infty}
dM
\,n(M,z)\,p(M^\mathrm{ob}|M)\;,
\label{eq:nln}
\end{eqnarray}
where $dV/dz$ is the cosmology-dependent comoving volume element per
unit redshift interval. The lower observed mass bin is bound by
$M^\mathrm{ob}_{\ell,m=0}=M_{\rm thr}(z)$, where $M_{\rm thr}(z)$ is
defined as the threshold value of the observed mass for a cluster to
be included in the survey (see Fig. \ref{f:selfuncphot}).  For the
halo mass function $n(M,z)$ in equation~(\ref{eq:nln}), we assume the
expression provided by \citet{tinker08}. Since the \eu\ selection
function has been computed for masses at $\Delta_c$ = 200 with respect
to the critical density, we use the \citet{tinker08} mass function
parameters relevant for an overdensity of $\Delta_{bk} = 200/\Omega_m(z)$
with respect to the background density. We note that in
equation~(\ref{eq:nln}) we have implicitly assumed that the survey sky
coverage $\Omega_\mathrm{sky}$ is independent of the observed mass,
which may not necessarily be the case if the sensitivity is not
constant over the survey area.

In equation (\ref{eq:nln}), $p(M^\mathrm{ob}|M)$ is the probability to
assign an observed mass $M^\mathrm{ob}$ to a galaxy cluster with true
mass $M$. Following \citet{lima05}, we use a lognormal probability
density, namely
\begin{equation}
p(M^\mathrm{ob}|M)=\frac{\exp[-x^2(M^\mathrm{ob})] }{ \sqrt{2\pi \sigma^2_{\ln
M}}}\,,
\label{eq:prob}
\end{equation}
where
\begin{equation}
x(M^\mathrm{ob})=\frac{\ln M^\mathrm{ob}-\ln M_\mathrm{bias}-\ln M}{\sqrt{2
\sigma^2_{\ln
M}}}\,.
\label{eq:m_mo}
\end{equation}
In the above equation $\ln M_\mathrm{bias}$ is the bias in the mass
estimation, which encodes any scaling relation between observable and
true mass and should not be confused with the bias in the galaxy
distribution. $\sigma_{\ln M}$
is the intrinsic scatter in the relation between true and observed
mass (see Section \ref{s:mod}). By inserting equation~(\ref{eq:prob}) into
equation~(\ref{eq:nln}), we obtain the expression for the cluster number
counts within a given mass and redshift bin,
\begin{eqnarray}
N_{\ell,m}& = & \frac{\Omega_\mathrm{sky}}{8\pi} \int_{z_\ell}^{z_{\ell+1}}dz\, 
\frac{dV}{dz} \nonumber \\ 
& & \int_0^{+\infty} dM\, n(M,z) \left[{\rm erfc}(x_m)-{\rm erfc}(x_{m+1})
\right]\,,
\label{eq:nln2}
\end{eqnarray}
where ${\rm erfc}(x)$ is the complementary error function and $x_m =
x(M^\mathrm{ob}_{l,m})$.

The FM for the averaged redshift-space cluster power spectrum within
the $\ell$-th redshift bin, the $m$-th wavenumber bin, and the $i$-th
angular bin can be written as
\begin{equation}
F_{\alpha \beta}^P=\frac{1}{8 \pi^2}\,\sum_{\ell,m,i} 
\frac{\partial \ln{\bar{P}(\mu_i,k_m,z_\ell)} }{ \partial p_\alpha} 
\frac{\partial \ln{\bar{P}(\mu_i,k_m,z_\ell)}}{\partial p_\beta}\,
V^\mathrm{eff}_{\ell,m,i} \, k_m^2\Delta k \Delta \mu
\label{eq:fm_pk}
\end{equation}
\citep[e.g.,][]{tegmark97,feldman94}, where the sums over $\ell$, $m$,
$i$ run over bins in redshift, wavenumber, and cosine of the angle
between $\boldsymbol k$ and the line of sight direction,
respectively. The quantity $V^{\mathrm{eff}}(\mu_i,k_m,z_\ell)$ represents the
effective volume accessible to the survey at redshift $z_{\ell}$ and
wavenumber $\boldsymbol k$ \citep{tegmark97,sartoris10}, and reads
\begin{equation}
V^\mathrm{eff}(\mu_i,k_m,z_\ell)=V_0(z_\ell)\frac{\tilde
n(z_\ell)\bar{P}(\mu_i,k_m,z_\ell)}{1+\tilde
n(z_\ell)\bar{P}(\mu_i,k_m,z_\ell)}\;.
\label{eq:veff}
\end{equation}
In the above equation, $V_0(z_\ell)$ is the total comoving volume
contained in the unity redshift interval around $z_\ell$, while $\tilde n
(z_\ell)$ is the
average number density of objects included in the survey at redshift $z_{\ell}$,
\begin{equation}
\tilde{n}(z_\ell) = \int_0^{+\infty} dM\, n(M,z_\ell)\,{\rm erfc}\left\lbrace
x[M_{\rm
thr}(z_\ell)\right\rbrace\;.
\end{equation}
The cluster power spectrum averaged over the redshift bin, appearing
in equation (\ref{eq:fm_pk}), can be written as
\begin{equation}
\bar{P}(\mu_i,k_m,z_\ell) = \frac{1}{S_\ell}\int^{z_{\ell+1}}_{z_\ell} dz
  \,\frac{dV}{dz} \,\tilde{n}^2(z)\, \tilde{P}(\mu_i,k_m,z)\;,
  \label{eq:barpk}
\end{equation}
where the normalisation factor $S_\ell$ reads
\begin{equation}
  S_\ell = \int^{z_{\ell+1}}_{z_\ell} dz
  \,\frac{dV}{dz}\, \tilde{n}^2(z)\;.
\end{equation}

\citet{sartoris12} pointed out the importance of taking into
account the contribution of cluster redshift space distortions for
constraining cosmological parameters. Following \citet{kaiser87}, we
calculate the redshift-space cluster power spectrum
$\tilde{P}(\mu_i,k_m,z_\ell)$ in the linear regime according to
\begin{equation}
\tilde{P}(\mu_i,k_m,z_\ell) = \left[ b_\mathrm{eff}(z_\ell) + f(z_\ell)
\mu^2\right]^2
P_\mathrm{L}\left(k_m,z_\ell\right)\,,
\label{eq:kai}
\end{equation}
where the power spectrum acquires a dependence on the cosine $\mu$ of
the angle between the wavevector $\boldsymbol k$ and the line-of-sight
direction. In the above equation, $b_\mathrm{eff}(z_\ell)$ is the linear
bias weighted by the mass function \citep[see equation 20
in][]{sartoris10},
\begin{equation}
b_\mathrm{eff}(z_\ell) = \frac{1}{\tilde n(z_\ell)}\int_0^{+\infty}
\mathrm{d}M\,n(M,z_\ell)\,{\rm erfc}\left\lbrace x[M_{\rm
thr}(z_\ell)]\right\rbrace\,
b(M,z_\ell)\,.
\label{eq:beff}
\end{equation}
The function $f(a) =d{\ln D(a)}/d{\ln a}$ is the logarithmic
derivative of the linear growth rate of density perturbations,
$D(a)$, with respect to the expansion factor $a$. 
$P_\mathrm{L}(k_m,z_\ell)$
is the linear matter power spectrum in real space, that we calculate
using the CLASS code \citep{class}. For the DM halo bias $b(M,z)$ we use the
expression provided by \cite{tinker10}.

Both the power spectrum and the number
counts FMs (equations \ref{eq:fm_nc} and \ref{eq:fm_pk}) are
computed in the redshift range defined by the \eu\ photometric
selection function shown in Fig. \ref{f:selfuncphot}, namely
$0.2\leq z \leq 2$, with redshift bins of constant width $\Delta z
= 0.1$. 
We note that the limiting precision with which the redshift $z_c$ of a
cluster is determined in the photometric survey is given by
$0.05(1+z_c)/N_{500,c}^{1/2}$, where $N_{500,c}$ is the total number of galaxies
assigned to the cluster. Therefore,
the bin width is always larger than the largest error on redshift
expected from the \eu\ photometric survey (see Section
\ref{s:clsel}). In equation (\ref{eq:fm_nc}), the observed mass range
extends from the lowest mass limit determined by the photometric
selection function ($M_\mathrm{thr}(z)$, see Fig.
\ref{f:selfuncphot}) up to $\mathrm{log}(M_\mathrm{ob}/M_{\odot})\leq
16$, with $\Delta \mathrm{log}(M_\mathrm{ob}/M_\odot) = 0.2$. In the
computation of the power spectrum FM (equation \ref{eq:fm_pk}), we adopt
$k_\mathrm{max}=0.14\,\mathrm{Mpc}^{-1}$, with $\Delta
\mathrm{log}(k\,\mathrm{Mpc}) = 0.1$. Finally, the cosine of the angle
between $\boldsymbol k$ and the line of sight direction, $\mu$, runs
in the range $-1\leq \mu \leq 1$ with $9$ equally spaced bins
\citep[see][]{sartoris12}.

\section{Cosmological and nuisance parameters}
\label{s:mod}
In this Section we discuss the cosmological parameters that have been
included in the FM analysis in order to predict the constraining power
of the \eu\ photometric cluster survey and we describe the peculiarity
of all the analysed models.  As a starting point, we consider all
the standard cosmological parameters for the concordance $\Lambda$CDM
model, whose fiducial values are chosen by following
\citet{planck13_16}: $\Omega_{\mathrm{m}}=0.32$ for the present-day
total matter density parameter, $\sigma_8=0.83$ for the normalisation of the
linear power spectrum of density perturbations,
$\Omega_{\mathrm{b}}=0.049$ for the baryon density parameter,
$H_0=67\,{\rm km\,s}^{-1} {\rm Mpc}^{-1}$ for the Hubble constant, and
$n_\mathrm{S}=0.96$ for the primordial scalar spectral index. We also
allow for a variation of the curvature parameter, whose fiducial
value $\Omega_{k}=0$ corresponds to spatial flatness.

\subsection{Model with dynamical Dark Energy}
In addition to the $\Lambda$CDM parameters, we also include parameters
describing a dynamical evolution of the DE component. In the
literature there are a number of models, characterised by different
parametrisation of the DE Equation of State (EoS henceforth)
evolution \citep[e.g.,][]{wetterich04}. In this paper we study the
parametrisation originally proposed by \citet{chevallier01}
and \citet{linder03} and then adopted in the DETF. We label this
parametrisation as the \cpl\
\de\ model, according to which the DE EoS can be written as
\begin{equation}
  w(a)\,=\,w_0\,+\,w_a(1-a)\;.
\label{eq:eos}
\end{equation}
We use $w_0=-1$ and $w_a=0$ as reference values for the two model parameters.
Thus, the cosmological parameter vector that we use in this first part of our
FM analysis reads
\begin{equation}
\boldsymbol{p} = \left\lbrace \Omega_{\mathrm{m}}, \sigma_8, w_0,w_a,
\Omega_{k}, \Omega_{\mathrm{b}},
H_0,  n_\mathrm{S} \right\rbrace \,.
\label{eq:cpl_par}
\end{equation}
The constraints on the DE dynamical evolution obtained by
combining \emph{Planck} CMB data with WMAP polarisation and with LSS
information \citep{planck13_16}, are $w_0=-1.04^{+0.72}_{-0.69}$ and
$w_a\lesssim 1.3$ ($95$ per cent c.l.) assuming
$\Omega_k=0$. Currently, the evolution of the cluster number counts
alone does not constrain the DE equation of state
parameters. However, \citet{mantz14} were able to obtain: $w_0 =
-1.03 \pm 0.18$ and $w_a =-0.1^{+0.6}_{-0.7}$ (assuming $\Omega_k=0$),
by using CMB power spectra (1-year Planck data, SPT, ACT), SNIa, and
BAO data at different redshifts \citep[plus WMAP polarisation;][]{planck13_16}.

Despite these weak constraints on the CPL DE parametrisation
\citep{vikhlinin09c}, cluster counts are powerful probes of
the amplitude of the matter power spectrum.  For instance, $\sigma_8$
is constrained at the level of $\sim 8$ per cent both with optically
selected SDSS clusters \citep{RO10.1}, and with SZ selected SPT
clusters \citep{BE13.1}.  Moreover, clusters help breaking the
degeneracy between $\sigma_8$ and $\Omega_{\mathrm{m}}$ in CMB
datasets, improving the constraints on the amplitude of the matter
power spectrum by a factor of $\sim 2$ with respect to CMB constraints
alone \citep{RO10.1}.
\newline 

\subsection{Model with primordial Non-Gaussianity}
We extend the standard cosmological model by allowing
primordial density fluctuations to follow a non-Gaussian distribution
\citep[e.g.,][]{bartolo04,desjacques10,wang13}. When this happens, the
distribution of primordial fluctuations in Bardeen's
gauge-invariant potential $\Phi$ cannot be fully described by a power
spectrum - commonly parametrised by a power-law,
$P_\Phi(\boldsymbol{k}) = Ak^{n_\mathrm{S}-4}$ (where $k \equiv
\|\boldsymbol{k}\|$) - rather we need higher-order statistics such as
the bispectrum
$B_\mathrm{\Phi}(\boldsymbol{k}_1,\boldsymbol{k}_2,\boldsymbol{k}_3)$. Different
models of inflation are known to produce different shapes of this
bispectrum. Here we consider only the so-called \textit{local shape}, where
the bispectrum strength is maximised for \emph{squeezed}
configurations, in which one of the three momenta $\boldsymbol k_j$ is
much smaller than the other two.

Within the local shape scenario, we adopt the commonly used way to
parametrise the primordial non-Gaussianity, which allows us to write
Bardeen's gauge invariant potential as the sum of a linear
Gaussian term and a non-linear second-order term that encapsulates the
deviation from Gaussianity \citep[e.g.,][]{salopek90,komatsu01}:
\begin{equation}
\Phi = \Phi_\mathrm{G} + f_\mathrm{NL}\left( \Phi_\mathrm{G}^2 - \langle
\Phi_\mathrm{G}^2 \rangle \right)\;,
\label{eq:ng}
\end{equation}
where the free dimensionless parameter $f_\mathrm{NL}$ parametrises
the deviation from the standard Gaussian scenario. We stress that
there is some ambiguity in the normalisation of equation (\ref{eq:ng}). We
adopt the LSS convention (as opposed to the CMB convention, see
\citealt{pillepich10,grossi07,carbone08}) where $\Phi$ is linearly
extrapolated at $z = 0$ for defining the parameter
$f_\mathrm{NL}$. The relation between the two normalisations is
$f_\mathrm{NL} = D(z=\infty) (1+z) f_\mathrm{NL}^\mathrm{CMB}/D(z=0) \simeq 1.3
f_\mathrm{NL}^\mathrm{CMB}$, where $D(z)$ is the linear growth
factor with respect to the Einstein-de Sitter
cosmology.

If the density perturbation field is non-Gaussian and has a positively
(negatively) skewed distribution, the
probability of forming large overdensities - and thus large collapsed structures
- is enhanced (suppressed). Thus, the shape and the evolution of the mass
function of DM halos change \citep[e.g.,][]{matarrese00, grossi09,loverde08}.
Following the 
prescription by \citet{loverde08} one can modify the mass function $n(M,z)$ in
equation (\ref{eq:nln}) to
take into account the \nong\ correction as follows
\begin{equation}
\label{eq:MF_fin}
n(M,z) = n^\mathrm{(G)}(M,z)
\frac{n_\mathrm{PS}(M,z)}{n^\mathrm{(G)}_\mathrm{PS}(M,z)}\;.
\end{equation}
In the previous equation,
$n^{\mathrm{(G)}}(M,z)$ is the mass function in the reference Gaussian
model, while $n_\mathrm{PS}(M,z)$ and
$n^\mathrm{(G)}_\mathrm{PS}(M,z)$ represent the \citet{press74} mass
functions in the \nong\ and reference Gaussian models, respectively
\citep[see the full equations in][]{sartoris10}.

In \nong\ scenarios the large-scale clustering of DM halos also changes. This
modification is quite
important because it alters in a fairly unique way the spatial distribution of
tracers of the cosmic structure, including galaxy clusters
\citep{dalal08, matarrese08, giannantonio10}. Specifically, the linear bias
acquires an extra scale dependence due to primordial non-Gaussianity, and can be
written as \citep{matarrese08}
\begin{equation}
b(M,z,k) = b^\mathrm{(G)}(M,z) + \left[ b^\mathrm{(G)}(M,z)-1 \right]
\delta_\mathrm{c}(z) \Gamma_R(k)\;,
\end{equation}
where $\Gamma_R(k)$ encapsulates the dependence on the scale and is
given by an integral over the primordial bispectrum.

To summarise, the cosmological parameter vector in this non-Gaussian
extension of the $\Lambda$CDM model is
\begin{equation}
\boldsymbol{p} = \left\lbrace \Omega_{\mathrm{m}}, \sigma_8, w_0,w_a,
\Omega_{k}, \Omega_{\mathrm{b}},
H_0,  n_\mathrm{S}, f_\mathrm{NL} \right\rbrace\;.
\label{eq:ng_par}
\end{equation}
We assume $f_\mathrm{NL}=0$ as the fiducial value of the non-Gaussian
amplitude.

The level of primordial non-Gaussianity has recently been constrained
to high precision thanks to \emph{Planck} CMB data
\citep{PL13.2}, $-4 \lesssim f_\mathrm{NL}\lesssim 11$, for 
the case of a local bispectrum shape\footnote{The \emph{Planck} CMB constraints
on primordial
  non-Gaussianity have been converted here into the LSS convention.}.
Bounds from galaxy cluster
abundance show consistency with the Gaussian scenario, $-91\lesssim
f_\mathrm{NL}\lesssim 78$
\citep{shandera13}. 
Constraints from the distribution of clusters
  are even less stringent \citep{mana13}. The clustering of \eu\ spectroscopic
galaxies
alone is expected to restrict the allowed non-Gaussian parameter space
down to $\Delta f_\mathrm{NL} \sim $ a few
\citep{CA08.1,VE09.1,FE11.1}. 
\newline

\subsection{Parametrise deviation from General Relativity}
We studied another extension to the standard $\Lambda$CDM cosmology,
based on deviations of the law of gravity from GR. As a matter
of fact, a number of non-standard gravity models have been proposed in the
literature \citep[e.g.,][]{hu_sawicki07,capozziello11,amendola12} in order to
explain the
low-redshift accelerated expansion of the Universe without need for
the DE fluid. Many of these models give rise to modifications of the
late-time linear growth of cosmological structure, which can be parametrised as
\begin{equation}
 \frac{d\ln D(a)}{ d\ln a} = \Omega_{\rm m}^\gamma(a)\;,
\label{eq:gamma}
\end{equation}
where $\gamma$ is dubbed the \emph{growth
index} \citep[e.g.][]{lahav91}. GR predicts a nearly constant and
scale-independent value of $\gamma \simeq 0.55$ \citep[e.g.][]{linder05}. Significant
deviations from this value would hence signal a violation of the
standard theory of gravity on large scales. The corresponding vector
of cosmological parameters in this case reads
\begin{equation}
\boldsymbol p = \left\lbrace \Omega_{\mathrm{m}}, \sigma_8, w_0,w_a,
\Omega_{k}, \Omega_{\mathrm{b}},
H_0,  n_\mathrm{S}, \gamma \right\rbrace\;,
\label{eq:g_par}
\end{equation}
with $\gamma=0.55$ taken as our reference value. Using number counts
of X-ray clusters alone, \citet{mantz14b} have found values of
$\gamma$ consistent with GR ($\gamma = 0.48\pm0.19$). From a sample of
SZ-selected clusters in SPT survey $\gamma=0.73\pm0.28$ has been
found \citep{bocquet14}. \citet{LO12.1} have directly
constrained the $f(R)$ model by \citet{hu_sawicki07} by exploiting an
optically selected cluster sample.\newline

\subsection{Model with non-minimal neutrino mass}
In our analysis we also consider the case of massive
neutrinos, 
with the associated density parameter $\Omega_{\nu}$ as the relevant
parameter to be constrained.  $\Omega_{\nu}$ is related to the total
neutrino mass, $\sum_{i}^{N_{\nu}}
   m_{\nu,i}$, through the relation:
\begin{equation}
 \Omega_{\nu} =\frac{\rho_{\nu}}{\rho_c} = \frac{\sum_{i}^{N_{\nu}}
   m_{\nu,i}}{93.14\,\mathrm{h^2\, eV}},
\end{equation}
where $\rho_{\nu}$ and $\rho_c$ are the $z=0$ neutrino and critical mass
densities, respectively, and $N_{\nu}$ is the number of massive neutrinos.
A larger value of $\Omega_{\nu}$ acts on the observed matter power
spectrum in two ways \citep[e.g.][]{lesgourgues06,marulli11,massara14}.
The peak of the power spectrum is shifted to larger scale, because a
larger value of the radiation density postpones the time of equality.
Moreover, since neutrinos free-stream over the scale of galaxy clusters, they do
not contribute to the clustered collapsed mass on such a scale. As a
consequence, the halo mass function at fixed value of $\Omega_{m}$
will be below the one expected in a
purely CDM model.
\cite{Brandbyge10} have shown that results from N -body
simulations with massive neutrinos can be reproduced in a more
accurate way by using the \citet{tinker08} halo mass function with
\begin{equation}
\rho_m \rightarrow \rho_{CDM} + \rho_b = \rho_m - \rho_\nu\,,
\end{equation}
where $\rho_m, \rho_{CDM}, \rho_b$ and $\rho_\nu$ are the total mass, CDM, and baryon and neutrino
densities.  Based on the analysis of an extended set of
N--body simulations, \citet{castorina14} and \citet{costanzi13b} have
shown that, since neutrinos play a negligible role in the
gravitational collapse, only the contribution of cold dark matter and
baryons to the power spectrum has to be used to compute the
r.m.s. of the linear matter perturbations, $\sigma(R)$, in the
computation of the halo mass function and linear bias:
\begin{equation}
P_m \rightarrow P_{CDM} (k) = P_m (k) \left[\frac{\Omega_{CDM} T_{CDM} (k,z) +
\Omega_b T_b (k,z)} { (\Omega_b+\Omega_{CDM}) T_m(k,z)}\right]^2\,.
\end{equation}
Here $T_{CDM}, T_b$ and $T_m$ are the CDM, baryon, and total matter
transfer functions respectively, and $P_m$ is the total matter power
spectrum.

Hence, the cosmological parameter vector we use in this case is:
\begin{equation}
\boldsymbol p = \left\lbrace \Omega_{\mathrm{m}}, \sigma_8, w_0,w_a,
\Omega_{k}, \Omega_{\mathrm{b}},
H_0,  n_\mathrm{S}, \Omega_{\nu} \right\rbrace\;,
\label{eq:nu_par}
\end{equation}
with a fiducial value of $\Omega_{\nu} = 0.0016$ that corresponds to $\sum
m_{\nu} =
0.06$ for three degenerate neutrinos \citep{carbone12,mantz14b}.  Currently,
great attention has been devoted to derive
constraints on the neutrino mass from the combination of galaxy
clusters with other LSS observables. The analysis of the Planck SZ
cluster sample resulted in $\sum m_{\nu} = 0.20 \pm 0.09$ eV
\citep{planck13_20}. \citet{mantz14}, combining
cluster, CMB,  SN1a and BAO data,  found $\sum m_{\nu}< 0.38$ eV at
95.4 per cent c.l. in a wCDM universe.  \citet{costanzi14}
found $\sum m_{\nu} <0.15$ eV (68 per
cent c.l.) in a $\Lambda$CDM universe,
for a three active neutrino scenario, using cluster counts, CMB,
BAO, Lyman-$\alpha$, and cosmic shear data.  In \citet{bocquet14} the
analysis of SPT cluster sample resulted in $\sum m_{\nu} = 0.148 \pm 0.081$
eV.\newline

\begin{figure}
\includegraphics[width=0.5\textwidth]{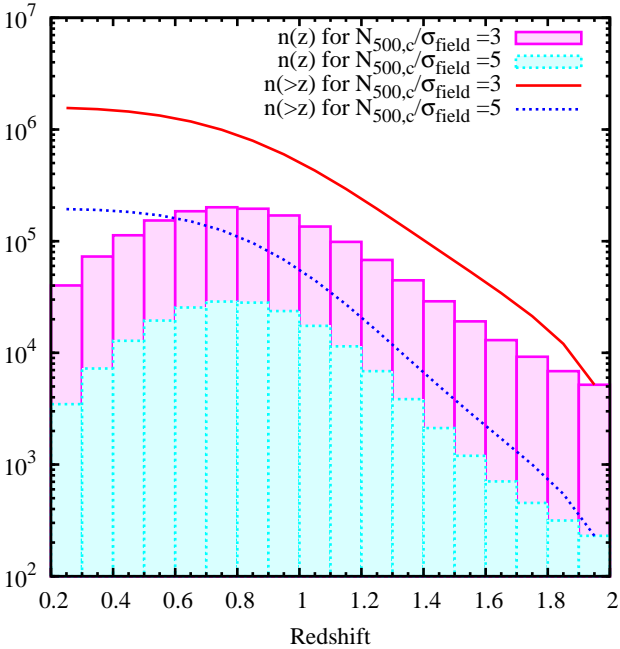}
\caption{Number of clusters above a given redshift to be detected
  with overdensities $N_{500,c}/\sigma_{\mathrm field}>5$ and $>3$  in the
  \eu\ photometric survey (dotted blue and solid red lines,
  respectively). We also show the number density of clusters expected
  to be detected within redshift bins of width $\Delta z=0.1$ for the
  same detection thresholds (dotted cyan and solid magenta histograms,
  respectively).The numbers have been obtained by using the reference
values of cosmological and nuisance parameters (see Section \ref{s:mod}).}
\label{fig:nz_eu}
\end{figure}

\subsection{Parameters of the mass--observable scaling relation}
\label{s:sr}
In our FM analysis, besides the cosmological parameter
vectors detailed above, we also include four extra parameters to model
intrinsic scatter and bias
in the scaling relation between the observed and true galaxy
cluster masses (see equation \ref{eq:m_mo} above). We assume the following
parametrisation for the bias and the scatter, respectively:
\begin{equation}
\ln M_\mathrm{bias}(z) = B_{M,0} + \alpha\, \ln\,(1+z)
\nonumber
\end{equation}
and 
\begin{equation}
\sigma_{\ln M}^2 (z) = \sigma_{\ln M,0}^2 - 1 + (1+z)^{2\beta} \;.
\label{eq:nuis}
\end{equation}
We select the following fiducial values 
\begin{equation}
 \boldsymbol{p}_\mathrm{nuisance,\,F}=\left\lbrace  B_{M,0} =0 , \alpha = 0
,\sigma_{\ln M,0} = 0.2,  \beta =0.125 \right\rbrace\;.
\label{eq:mass_par}
\end{equation}
We refer to these four parameters as \emph{nuisance} parameters
henceforth. With the fiducial nuisance parameter vector there is no
bias in the true mass-observable relation and the value of the scatter
at $z = 0$ is in accordance with \citet{rykoff12}. Also, the
fiducial value for $\beta$ makes the scatter increase
with redshift, reaching $\sigma_{\ln M} \simeq 0.6$ at the maximum
redshift of the \eu\; survey ($z_\mathrm{max} =2$). 

Currently, the mass-observable relations are not known over the full
redshift range that will be covered by \eu. In the \eu\ survey it will
be possible to calibrate such relation with its uncertainties thanks
to the weak lensing and spectroscopic surveys. We estimate that
\eu\ has the potential to calibrate the scaling relation to
$\leq 15$ per cent accuracy out to $z \leq 1.5$ (see Appendix~\ref{app:mass}).

In the following Section, we will consider the two extreme cases where
we assume (i) no prior information on the nuisance parameters, and
(ii) perfect knowledge of the mass-observable relation.

\section{Results}
\label{s:res}

Here, we present the constraints on the cosmological parameter vectors
introduced in the previous Section, using the FM formalism. As a first
result, we plot in Fig. \ref{fig:nz_eu} the histograms corresponding
to the redshift distributions, $n(z)=\Delta z \,dN/dz$ (equation\ref{eq:nln2}),
of \eu\ photometric galaxy clusters, obtained by adopting the two
selection functions, which correspond to the two different detection
thresholds $N_{500,c}/\sigma_{\mathrm field}> 3$ and 5 (see
Fig. \ref{f:selfuncphot}), and by using the reference values of cosmological
and nuisance parameters.
The two curves show the corresponding
cumulative redshift distributions, $n(>z)$, i.e., the total number of clusters
detected above a given redshift. \eu\ will detect $\sim 2 \times10^5$
objects with $N_{500,c}/\sigma_\mathrm{field}
\ge 5$ at all redshifts, with about $\sim 4 \times 10^4$ of them at
$z\ge 1$. By lowering the detection threshold down to
$N_{500,c}/\sigma_\mathrm{field} =3$, these numbers rise up to $\sim 2
\times10^6$ clusters at all redshifts, with $\sim 4 \times 10^5$ of them at
$z\ge 1$. The large statistics of clusters at $z\ge 1$ provides a wide
redshift leverage over which to follow the growth rate of
perturbations. 
As a comparison,
DES will detect $\sim 1.7 \times 10^{5}$ clusters (with more than 10
bright red-sequence galaxies)
and with masses greater than $\sim 5\times 10^{13}\, M_\odot$ out to $z\sim1.5$
in the survey area of 5000 deg$^2$
\footnote{https://www.darkenergysurvey.org/reports/proposal-standalone.pdf}
eROSITA \citep{pillepich12} will detect $\sim 9.3 \times 10^{4}$
clusters with masses greater than $\sim 5\times 10^{13}\, M_\odot$ in the
survey area of 27.000 deg$^2$, almost all at $z<1$.

\begin{figure*}
\includegraphics[width=0.47\textwidth]{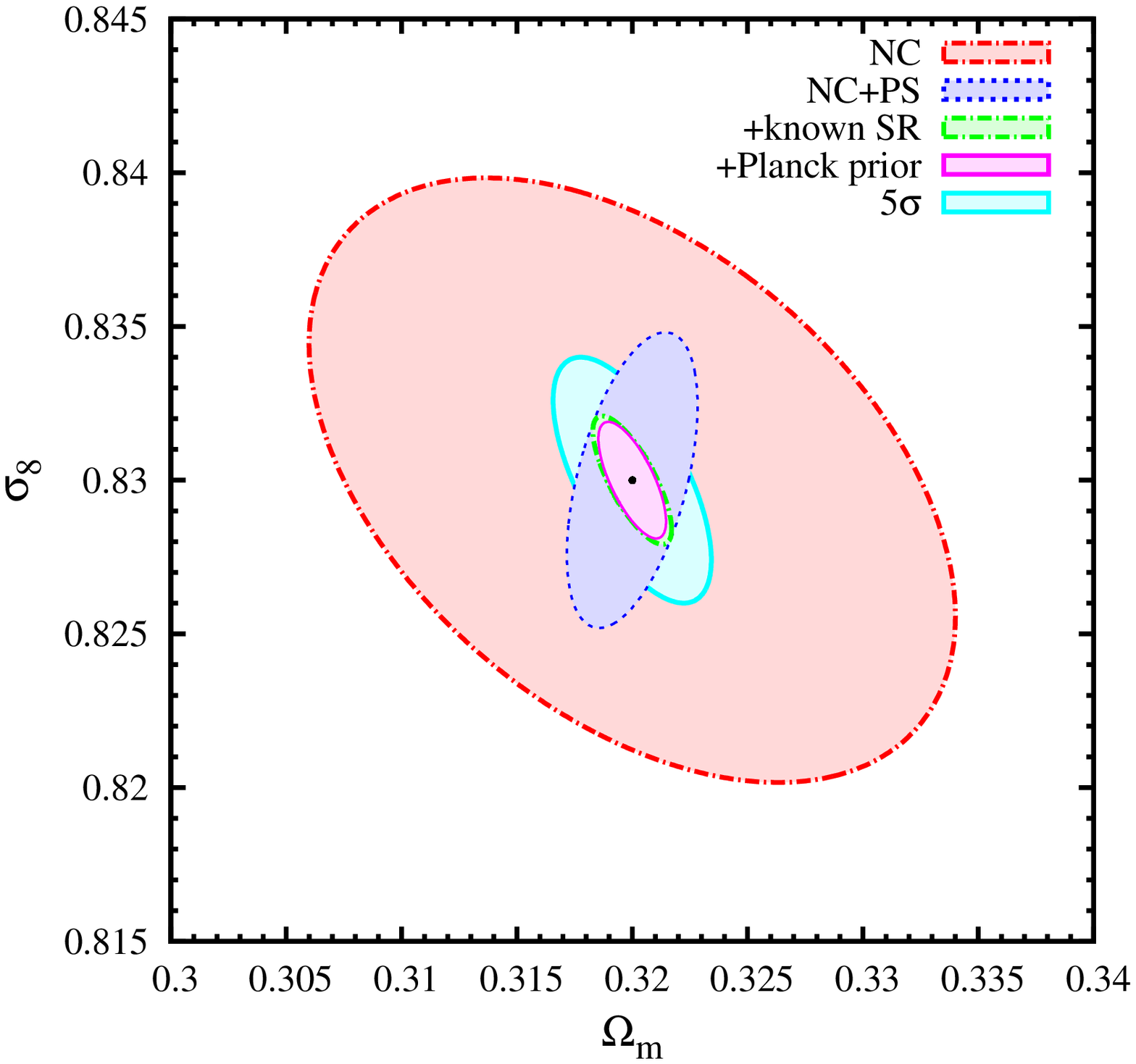}
\includegraphics[width=0.47\textwidth]{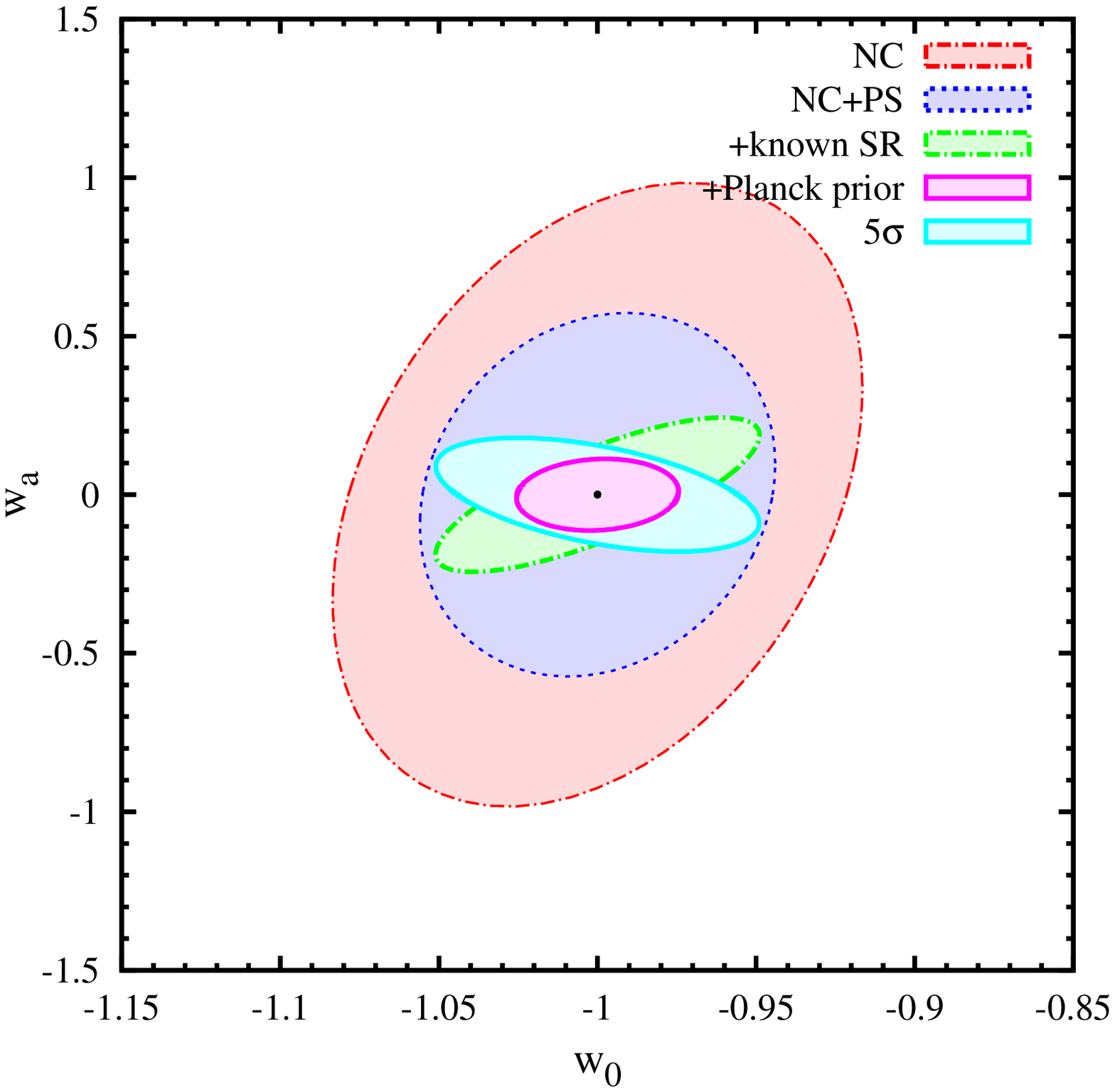}
\caption{Constraints at the $68$ per cent c.l. on the parameters
$\Omega_{\mathrm{m}}$ and $\sigma_8$ (left panel) and on the parameters $w_0$
and $w_a$ for the DE EoS evolution (right panel). In each panel, we show
forecasts for the $N_{500,c}/\sigma_\mathrm{field}\ge 3$
\eu\ photometric cluster selection obtained by
(i) NC, the FM number counts (red dash-dotted contours), (ii) NC+PS,
the combination of FM NC and power spectrum (PS) information (blue
dotted contours), (iii) NC+PS+known SR, i.e. by additionally assuming a perfect
knowledge of the nuisance parameters (green dash-dotted contours), and
(iv) NC+PS+known~SR+\emph{Planck}~prior, i.e. by also adding information
from
\emph{Planck} CMB data (magenta solid contours). With cyan solid lines we show
forecasts for
the $N_{500,c}/\sigma_\mathrm{field}\ge 5$ \eu\ photometric cluster selection
in the case NC+PS+known~SR+\emph{Planck}~prior (labelled $5 \sigma$).
Planck information includes
prior on $\Lambda$CDM parameters and the DE EoS parameters.}
\label{fig:constrde}
\end{figure*}

In
 Figs.~\ref{fig:constrde}, \ref{fig:constrng}, \ref{fig:constrgamma},
 and \ref{fig:constrnu} we show the forecasted constraints from \eu\
 photometric clusters on suitable pairs of cosmological
 parameters. The ellipses in these figures always correspond to the
 $68$ per cent c.l. after marginalisation over all other cosmological
 parameters and nuisance parameters. In each of these figures, the
 blue dotted contours are obtained by combining the number counts (NC)
 FM (equation \ref{eq:fm_nc}) and the cluster power spectrum (PS) FM
 (equation\ref{eq:fm_pk}), assuming no prior information on any of the
 cosmological and nuisance parameters. Also, the cluster sample is
 defined by the selection $N_{500,c}/\sigma_\mathrm{field} \ge 3$. The
 green dash-dotted contours are obtained in the same way except for
 the addition of strong priors on the nuisance parameters,
 i.e. assuming perfect knowledge of the scaling relation between the
 true and the observed cluster mass (this is labelled as "+known SR" in
 the figures). The magenta solid contours have been obtained by
 further introducing prior information from \emph{Planck} data
 (label-ed "+\emph{Planck} prior" in the figures). Finally, the cyan
 solid contours represent the same combination of information as the
 magenta solid ones (NC+PS+known SR+ \emph{Planck} prior) obtained
 from the cluster sample with selection corresponding to
 $N_{500,c}/\sigma_\mathrm{field} \ge 5$.  In the figures, we indicate
 these contours with the label $5\sigma$.

When using the \emph{Planck} priors, we take for the \cpl\ DE model
the correlation matrix obtained by combining \emph{Planck} CMB data
with the BAOs from \citet{planck13_16}\footnote{Available at \texttt{
http://wiki.cosmos.esa.int/\\planckpla/index.php/Cosmological\_Parameters}}
for the parameters of the $\Lambda$CDM cosmology (assuming $\Omega_k=0$), plus
$w_0$ and
$w_a$\footnote{PLA/base\_w\_wa/planck\_lowl\_lowLike\_BAO}. For the
non-Gaussian case, we use priors from the \emph{Planck} obtained
for the $\Lambda$CDM model plus $\Omega_{k,0}$
parameters\footnote{PLA/base\_omegak/planck\_lowl\_lowLike}. We also
added a flat prior on the level of non-Gaussianity corresponding to
$-5.8\le f^{\mathrm{CMB}}_\mathrm{NL}\le5.8$. Finally, for the modified gravity
and
the neutrino scenario we also used priors from the \emph{Planck}
analysis carried out over the parameters of the $\Lambda$CDM model plus $\Omega_k$.

In Fig. \ref{fig:constrde},
we show the constraints on $\Omega_{\mathrm{m}}$ and $\sigma_8$
(left panel), as well as those on the two CPL DE parameters $w_0$ and
$w_a$ (right panel). The contours on the
$\Omega_{\mathrm{m}}-\sigma_8$ plane for the combination of number
counts and clustering of $N_{500,c}/\sigma_\mathrm{field} \ge 3$ galaxy
clusters are rather tight. The information provided by the number
density of clusters alone defines the degeneracy direction between
$\Omega_{\mathrm{m}}$ and $\sigma_8$, with the following constraints:
$\Delta \Omega_m = 0.009,
\Delta \sigma_8 = 0.006$. Information from the cluster power spectrum
alone does not provide stringent constraints on the
$\Omega_{\mathrm{m}}-\sigma_8$ plane. However, using the combination
of the PS with NC FM, the values of both parameters are constrained to
high accuracy: $\Delta \Omega_m = 0.0019, \Delta \sigma_8 = 0.0032$
(see Table~\ref{tab:fom}).  By assuming a perfect knowledge of the
scaling relation between true and observed cluster mass, the bounds
improve significantly. This is especially true for $\sigma_8$, which
is more affected by the nuisance parameters than
$\Omega_{\mathrm{m}}$. Including information from the \emph{Planck}
priors does not improve the forecasted constraints significantly, in
keeping with the expectation that the
\eu\ cluster bounds are, by themselves, competitive with CMB
bounds.

Taking the $\Lambda$CDM as a reference model, its parameters will be 
constrained with a precision of $\sim 10^{-3}$,
\begin{equation}
 \Delta \Omega_m=5.9\, 10^{-4}, 
\Delta \sigma_8=4.9\, 10^{-4}, 
\Delta h=7.2\, 10^{-4},\nonumber
\end{equation}
\begin{equation}
\Delta \Omega_b=8.4\, 10^{-4},
\Delta n_s=3.3\,10^{-3}
\end{equation}
thanks to the unprecedented number of clusters that will be detected
at high redshift.  These constraints have been obtained with the
$N_{500,c}/\sigma_\mathrm{field}\ge 3$ selection function, from
cluster number counts and power spectrum, by assuming \textit{strong
prior} on the nuisance parameters, and no prior from Planck.

These results emphasise the importance of exploring the high-redshift
clusters in survey mode. Of course a good knowledge of the
astrophysical process taking place in clusters is fundamental to
calibrate the mass-observable scaling relations, and also to optimise
the detection algorithms.  Hence detailed follow-ups of restricted
samples of clusters
\citep[such as, e.g., CLASH, CCCP, WtG][]{postman14,rosati14,hoekstra12,linden14} retain a crucial importance.

On  the other hand, the inclusion of \emph{Planck}
priors shall bring a substantial improvement over the
bounds to the DE parameters. This result is expected, since the
CMB data provides stringent constraints on the curvature, thereby breaking the
degeneracy
between $\Omega_k$ and the evolution of the DE EoS
\citep{sartoris12}.  The
contribution of the PS information is less important for $(w_0,w_a)$
with respect to $(\Omega_{\mathrm{m}},\sigma_8)$: however, the
FoM increases from $\sim 30$ in case
of NC alone to $\sim 73$ for NC+PS constraints (see Table
\ref{tab:fom}). 
For both DE EoS parameters, it is crucial to have a well calibrated
scaling relation over the redshift range sampled by the cluster
survey \citep{sartoris12}. Indeed, by combining NC and
PS, and assuming perfect knowledge of the scaling
relation increases the FoM to $\simeq 291$. When we include the Planck data,
i.e. we set a prior on the curvature, we obtain FoM$=802$, with
$\Delta w_0 = 0.017 $ and $\Delta w_a = 0.074$ (see Table~\ref{tab:fom}).

When we restrict our analysis to the wCDM model (that is characterised
by the six free parameters $\Omega_m, \sigma_8, h, \Omega_b, n_s, w$),
we obtain $\Delta w = 0.005$. If we also add $w_a$ as a free
parameter, we obtain $\Delta w_0 = 0.013 $ and $\Delta w_a = 0.048$.
These constraints have been obtained with the
$N_{500,c}/\sigma_\mathrm{field}\ge 3$ selection function, from
cluster number counts and power spectrum, by assuming \textit{strong
prior} on the nuisance parameters, and no prior from Planck.

In both panels of Fig. \ref{fig:constrde}, the adoption of a more
conservative cluster selection ($N_{500,c}/\sigma_\mathrm{field} \ge 5$)
significantly worsens the forecasted cosmological constraints. For
instance, the FoM is degraded down to $209$ in the best-case scenario,
as a consequence of the significantly degraded statistics
corresponding to the higher selection threshold.

\begin{figure}
\includegraphics[width=0.47\textwidth]{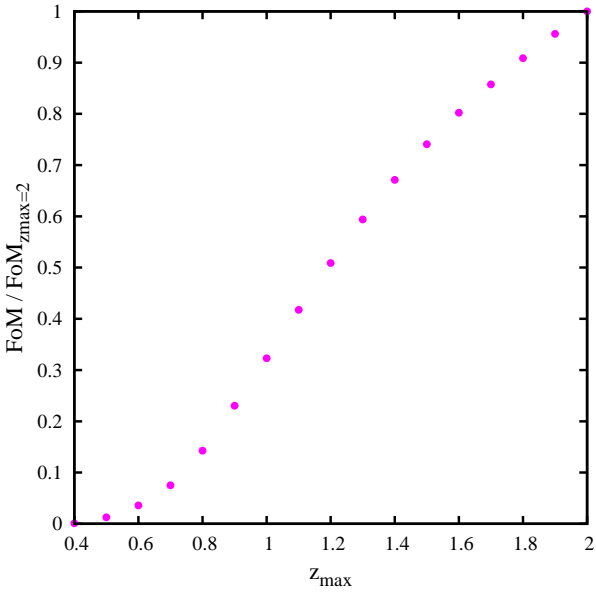}
\caption{Relative FoM for number counts in the
$N_{500,c}/\sigma_\mathrm{field}\ge 3$ \eu\ photometric cluster selection, as a
function of the limiting redshift
$\mathrm{z_{max}}$ of the survey, i.e. the ratio between the FoM
evaluated over $0.2 \leq z \leq z_{max}$ and the FoM evaluated over
$0.2 \leq z \leq 2.0$.}
\label{fig:fom}
\end{figure}

\begin{figure}
\includegraphics[width=0.47\textwidth]{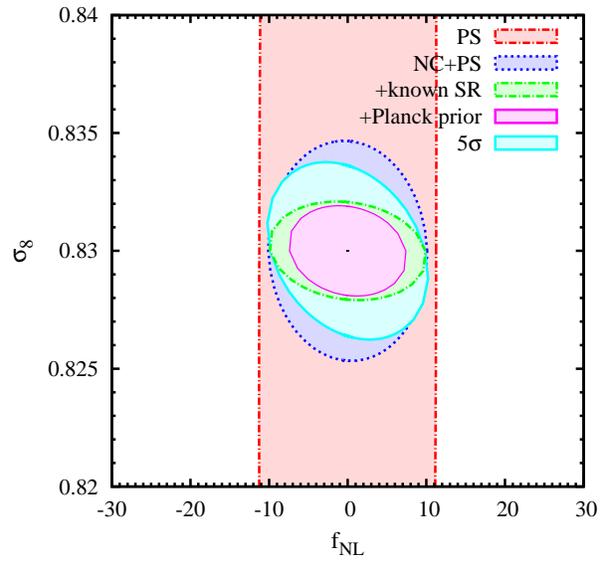}
\caption{Constraints at the $68$ per cent c.l. on the
  $f_\mathrm{NL}-\sigma_8$ parameters. We show
forecasts for the $N_{500,c}/\sigma_\mathrm{field}\ge 3$
\eu\ photometric cluster selection obtained by
(i) PS, the FM power spectrum (red dash-dotted contours), (ii) NC+PS,
the combination of FM number counts (NC) and PS information (blue
dotted contours), (iii) NC+PS+known SR, i.e. by additionally assuming a perfect
knowledge of the nuisance parameters (green dash-dotted contours), and
(iv) NC+PS+known~SR+\emph{Planck}~prior, i.e. by also adding information
from
\emph{Planck} CMB data (magenta solid contours). With cyan solid lines we show
forecasts for
the $N_{500,c}/\sigma_\mathrm{field}\ge 5$ \eu\ photometric cluster selection
in the case NC+PS+known~SR+\emph{Planck}~prior (labelled $5 \sigma$).
Planck
  information includes prior on $\Lambda$CDM+$\Omega_k$+$f_\mathrm{NL}$
  parameters.}
\label{fig:constrng}
\end{figure}

\begin{figure}
\includegraphics[width=0.47\textwidth]{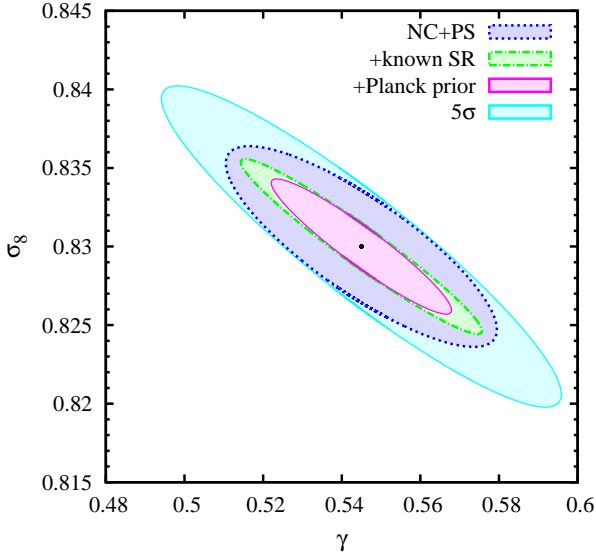}
\caption{Constraints at the $68$ per cent c.l. on the
  $\gamma-\sigma_8$ parameter plane. We show forecasts for the
$N_{500,c}/\sigma_\mathrm{field}\ge 3$
\eu\ photometric cluster selection obtained by
(i) NC+PS, the combination of FM number counts (NC) and power spectrum
(PS) information (blue dotted contours), (ii) NC+PS+known SR, i.e. by
additionally assuming a perfect knowledge of the nuisance parameters
(green dash-dotted contours), and (iii)
NC+PS+known~SR+\emph{Planck}~prior, i.e. by also adding information
from \emph{Planck} CMB data (magenta solid contours). With cyan solid
lines we show forecasts for the $N_{500,c}/\sigma_\mathrm{field}\ge
5$ \eu\ photometric cluster selection in the case
NC+PS+known~SR+\emph{Planck}~prior (labelled $5 \sigma$).  Planck
information includes prior on $\Lambda$CDM+$\Omega_k$
parameters.}
\label{fig:constrgamma}
\end{figure}

\begin{figure}
\includegraphics[width=0.47\textwidth]{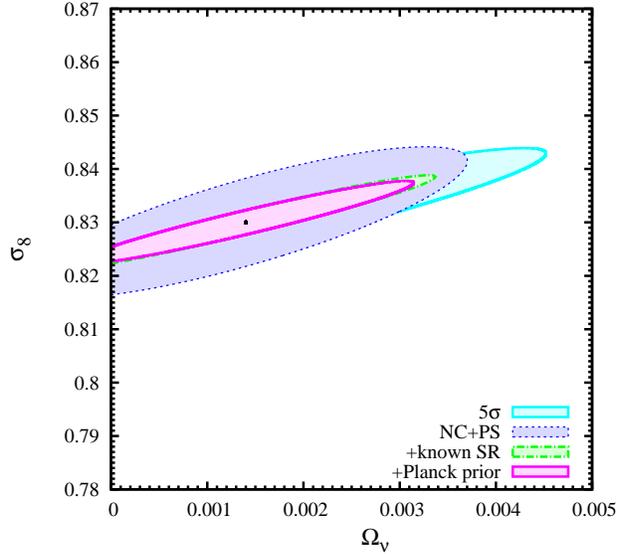}
\caption{Constraints at the $68$ per cent c.l. in the
  $\Omega_{\nu}-\sigma_8$ parameter plane. 
We show forecasts for the
$N_{500,c}/\sigma_\mathrm{field}\ge 3$
\eu\ photometric cluster selection obtained by
(i) NC+PS, the combination of FM number counts (NC) and power spectrum
(PS) information (blue dotted contours), (ii) NC+PS+known SR, i.e. by
additionally assuming a perfect knowledge of the nuisance parameters
(green dash-dotted contours), and (iii)
NC+PS+known~SR+\emph{Planck}~prior, i.e. by also adding information
from \emph{Planck} CMB data (magenta solid contours). With cyan solid
lines we show forecasts for the $N_{500,c}/\sigma_\mathrm{field}\ge
5$ \eu\ photometric cluster selection in the case
NC+PS+known~SR+\emph{Planck}~prior (labelled $5 \sigma$).  Planck
information includes prior on $\Lambda$CDM+$\Omega_k$ parameters.}
\label{fig:constrnu}
\end{figure}

In Fig.~\ref{fig:fom}, we show how the FoM depends on the limiting
redshift of the survey. The FoM shown in this figure refers
to number counts (NC)
in the $N_{500,c}/\sigma_\mathrm{field}\ge 3$
\eu\ photometric cluster selection.
The FoM for a survey reaching out to $z \leq 1.2$ is only half the FoM
of an equivalent survey reaching out to $z \leq 2$. It is therefore
important that the redshift range covered by the survey be large
enough to allow a comparison of the behaviour of DE over a sufficiently long
cosmological timescale.  In this sense, the \eu\ survey will have a
unique advantage over other existing and planned surveys.

\begin{table*}
\centering
 \caption{Figure of Merit (FoM) and constraints on cosmological parameters as
obtained by progressively adding 
   the FM information for different models, for two different detection
thresholds ($N_{500,c}/\sigma_\mathrm{field}\ge 3$ and 5).
Constraints  are shown at $68$ per cent c.l. after marginalisation over all
other
cosmological parameters and nuisance parameters in the arrays.}
 \begin{tabular}{|l|cccccccc|}
\hline
\multicolumn{8}{|c|}{$N_{500,c}/\sigma_\mathrm{field}\ge 3$ \eu\ photometric
cluster selection}\\
\hline
Parameter arrays:&\multicolumn{5}{|c|}{Eqs. \ref{eq:cpl_par} \&
\ref{eq:mass_par}} & Eqs. \ref{eq:g_par} \& \ref{eq:mass_par} & Eqs.
\ref{eq:ng_par} \& \ref{eq:mass_par} & Eqs. \ref{eq:nu_par} \&
\ref{eq:mass_par}\\
Constraints: & FoM &$\Delta w_0$ &  $\Delta w_a$ &$\Delta \Omega_m$ &  $\Delta
\sigma_8$ &$\Delta \gamma$ &  $\Delta f_{NL}$ & $\Delta \Omega_{\nu}$ \\ 
\hline
NC+PS& 73&0.037 &0.38 &0.0019&0.0032&0.023&6.67&0.0015\\
NC+PS+known SR & 291&0.034&0.16&0.0011&0.0014&0.020&6.58&0.0013\\
NC+PS+known SR+Planck &802&0.017&0.074&0.0010&0.0012&0.015&4.93&0.0012\\
\hline
\multicolumn{8}{|c|}{$N_{500,c}/\sigma_\mathrm{field}\ge 5$ \eu\ photometric
cluster selection}\\
\hline
NC+PS+known SR+Planck &209&0.034&0.12&0.0022&0.0026&0.034&6.74&0.0020\\
\hline
\hline
  \end{tabular}
\label{tab:fom}
 \end{table*}

 In Fig. \ref{fig:constrng}, we show cosmological constraints in the
 expanded parameter space which includes non-Gaussian primordial
 density fluctuations.  Specifically, we display the constraints in
 the $f_\mathrm{NL} - \sigma_8$ plane. Thanks to the peculiar
 scale-dependence that primordial non-Gaussianity induces on the
 linear bias parameter, the power spectrum of the cluster distribution
 turns out to be much more sensitive to $f_\mathrm{NL}$ than it is to
 $\sigma_8$. This is clearly demonstrated by the red dash-dotted
 contour, which shows forecasted constraints derived from cluster
 clustering alone. Quite clearly, $\sigma_8$ is basically
 unconstrained on the scale of the figure, while $f_\mathrm{NL}$ is
 constrained with an uncertainty $\Delta f_\mathrm{NL} \sim 7.4$. The
 addition of cluster number counts changes very little the bounds for
 primordial non-Gaussianity, however it improves substantially those
 for the amplitude of the matter power spectrum (see
 Table~\ref{tab:fom}). This helps to define the degeneracy between
 $f_\mathrm{NL}$ and $\sigma_8$ that are both related to the timing of
 structure formation.  Interestingly, the estimation of primordial
 non-Gaussianity is weakly sensitive to the nuisance parameters.
 Indeed, when a perfect knowledge of the scaling relation between true
 and observed cluster mass is assumed, only the constraints on
 $\sigma_8$ improve significantly. \emph{Planck} priors does not
 affect substantially the constraints on $f_\mathrm{NL}$. 

When we restrict our analysis to the $\Lambda$CDM model plus
the non-Gaussianity parameter
$f_\mathrm{NL}$, we obtain
$\Delta f_\mathrm{NL} = 6.44$.  This constraint has been obtained
with the $N_{500,c}/\sigma_\mathrm{field}\ge 3$ selection function,
from cluster number counts and power spectrum, by
assuming \textit{strong prior} on the nuisance parameters, and no
prior from Planck.  Forecast for
\emph{eROSITA} \citep{pillepich12} predict a similar precision,
since the narrower redshift range of this survey (with respect to \eu)
is compensated by its wider area, which allows a better sampling of
large scale modes.

We point out that in this analysis we are assuming the most commonly used
parametrisation of non-Gaussianity, where $f_\mathrm{NL}$ is considered
scale-invariant. However, there are models that predict otherwise. For these,
the combination of clusters and CMB data complement each other well, providing
tight constraints on the possible scale dependence of $f_\mathrm{NL}$.

As for the models including GR violation, we show in Fig.
\ref{fig:constrgamma} the constraints on $\sigma_8$ and the growth
parameter $\gamma$.  Similarly to the constraints on the
$\Omega_{\mathrm m}$--$\sigma_8$ plane, the constraints on $\gamma$
are not strongly affected by the inclusion of \emph{Planck} priors,
thus implying that galaxy clusters are by themselves excellent tools
to detect signature of modified gravity through its effect on the
growth of perturbations. Significant degradation of the
constraining power happens if a higher threshold for cluster
detection is chosen.

Restricting our analysis to the $\Lambda$CDM model 
plus the $\gamma$ parameter
we obtain $\Delta \gamma = 0.006$. 
This constraint has been obtained
with the $N_{500,c}/\sigma_\mathrm{field}\ge 3$ selection function,
from cluster number counts and power spectrum, by
assuming \textit{strong prior} on the nuisance parameters, and no
prior from Planck. 

Finally, we show in Fig. \ref{fig:constrnu} the joint cosmological
constraints on $\sigma_8$ and the neutrino density parameter
$\Omega_{\nu}$.  The presence of neutrinos with masses in the sub-eV
range requires higher values of $\sigma_8$: increasing $\Omega_{\nu}$
at fixed $\Omega_{\mathrm m}$ has the effect of shifting the epoch of
matter-radiation equality to a later time and to reduce the growth of
density perturbations at small scales in the post-recombination
epoch. As a consequence, a larger value of $\sigma_8$ is required to
compensate these effects. We use the Planck prior mainly to add
information on the geometry of the Universe, and the standard
$\Lambda$CDM parameters. We obtain the constraints $\Delta
\Omega_{\nu} =0.0012$ (corresponding to $\Delta \sum m_{\nu} = 0.01$ 
).  The constraints on the neutrino density parameter
are weakly affected by the inclusion of a prior on the nuisance
parameters. However, there is a degradation by a factor of $\sim 2$ of
the constraining power if the selection function with the higher
threshold for cluster detection is chosen (see Table~\ref{tab:fom}).

To gauge the impact of a particular choice of the selection function
on the cosmological constraints, we have so far shown our results for both
the $N_{500,c}/\sigma_\mathrm{field}\ge 3$ and the
$N_{500,c}/\sigma_\mathrm{field}\ge 5$
\eu\ photometric cluster selection functions.
As a further test, we consider the effect on the
$(w_0,w_a)$ constraints of adopting a flat selection function with
$\log (M_{200,c}) =13.9$, within $0.2\leq z \leq 2$. With
this \textit{flat} selection function there are less clusters than
with the $N_{500,c}/\sigma_\mathrm{field}\ge 3$ one, both in total
($\sim 1.4 \times 10^6$ vs. $\sim 1.6 \times 10^6$, respectively) and
within $0.4\lesssim z \lesssim 1.2$. However, the number of clusters
at $z>1$ is higher ($\sim 3.2 \times 10^5$) for the \textit{flat}
selection function than for the $N_{500,c}/\sigma_\mathrm{field}\ge 3$
one ($\sim 1.9 \times 10^5$).  The effect of a larger number of
high-$z$ clusters in the \textit{flat} selection function sample
compensates for the smaller total number of clusters in providing similar
constraints on the cosmological parameters
to those obtained with the $N_{500,c}/\sigma_\mathrm{field}\ge 3$
selection function sample (changes are $<10$\% on the constraints on the DE
parameters). This suggests that the precise shape
of the selection function has little impact on our results, and in any
case much less than its overall normalisation.

\section{Conclusions}
\label{s:concl}

In this paper, we presented a comprehensive analysis of the forecasts
on the parameters that describe different extensions of the standard
$\Lambda$CDM model. These were based on the selection
function of galaxy clusters from the wide photometric survey to be carried
out with the \eu\ satellite, a medium-size ESA mission to be launched
in 2020. We presented the derivation of this 
selection function and 
the Fisher Matrix formalism employed to derive cosmological constraints.
This is the same formalism that 
 has been used in the
\eu\ Red Book \citep{laureijs11}.  Our main results can be
summarised as follows.

\begin{itemize}
\item Using photometric selection, we found that \eu\ will detect
  galaxy clusters at $N_{500,c}/\sigma_\mathrm{field}\ge 3$ with a
  minimum mass of $\sim 0.9-1\times 10^{14}M_\odot$. 
  As a result, the \eu\ photometric cluster catalogue should
  include $\sim 2\times 10^{6}$ objects, with about one fifth of them
  at $z\ge 1$.
\item The \eu\ cluster catalogue has the potential of providing tight
  constraints on a number of cosmological parameters, such as the
  normalisation of the matter power spectrum $\sigma_8$, the total
  matter density parameter $\Omega_{\mathrm m}$, a redshift-dependent
  DE equation of state, primordial non-Gaussianity, modified gravity,
  and neutrino masses (see Table~\ref{tab:fom}).  We predict that most
  of these constraints will be even tighter than current bounds
  available from \emph{Planck}. The constraining power of the \eu\ cluster
catalogue
  relies on its unique broad redshift coverage, reaching out to $z=2$.
\item Knowledge of the scaling relation between the true and the 
observed cluster mass turns out to be one of the most important
factors determining the constraining power of the \eu\ cluster catalogue
for cosmology.  The \eu\ mission will have a distinct advantage in
this respect, namely the possibility to calibrate such relation, at
least up to $z=1.5$, with $\lesssim 10$ and $\lesssim 30$ per cent
accuracy, using the weak lensing and spectroscopic surveys,
respectively (see Appendix~\ref{app:mass}). The deep \eu\ survey will
allow to extend the calibration to even higher redshifts, although
with lower precision than in the wide survey, due to lower number
statistics.
\end{itemize}

With the future large surveys, like \eu, that will be carried out with
the next
generation of telescopes, the number of detected clusters from the
individual surveys will range from thousands to tens of thousands. As
we have shown in this paper, this will allow to constrain most
cosmological parameters to a precision level of a few per cent.
Currently, theoretical halo mass functions are defined with an
uncertainty of $\sim5$ per cent in the standard $\Lambda$CDM
model \citep[e.g.][]{tinker08}, and many efforts have been devoted in
the last years to better sample the high mass regime \citep{watson13}.
To maximally extract cosmological information from these cluster
surveys, it becomes critical to specify the theoretical halo mass
function to better than a few percent accuracy for a range of
cosmologies. A substantial effort is currently ongoing in this
direction \citep{grossi07,dalal08,cui12b,lombriser13,castorina14}.
Moreover, cosmological hydrodynamic simulations will have to precise
the impact of baryons on the shape of the mass profile, which has
already been shown to be quite
significant \citep{rudd08,stanek09,cui12,cui14}.

\section*{Acknowledgments}
We thank L. Pozzetti for providing us with her estimates of the number
densities of H$\alpha$-emitting galaxies in advance of publication. We
acknowledge useful discussions with O. Cucciati, S. Farrens,
A. Iovino, S. Mei, and F. Villaescusa.  We thank S. Andreon,
M. Brodwin, G. De Lucia, S. Ettori, M. Girardi, T. Kitching, G. Mamon,
J. Mohr, T. Reiprich for a careful reading of the manuscript.  BS
acknowledges financial support from MIUR PRIN2010-2011
(J91J12000450001) and a grant from ``Consorzio per la Fisica -
Trieste''.  BS and SB acknowledge financial support from the PRIN-MIUR
201278X4FL grant, from a PRIN-INAF/2012 Grant, from the ``InDark''
INFN Grant and from the ``Consorzio per la Fisica di Trieste''.  CF
has received funding from the European Commission Seventh Framework
Programme (FP7/2007-2013) under grant agreement n. 267251.  LM
acknowledges financial contributions from contracts ASI/INAF
n.I/023/12/0.  JW acknowledges support from the Transregional
Collaborative Research Centre TRR 33 - 'The Dark Universe'.  The
authors acknowledge the
\eu\ Collaboration, the European Space Agency and the support of a
number of agencies and institutes that have supported the development
of \eu. 

A detailed complete list is available on the \eu\ web
site (http://www.euclid-ec.org). In particular the Agenzia Spaziale
Italiana, the Centre National d'\'Etudes Spatiales, the Deutches
Zentrum fur Luft- and Raumfahrt, the Danish Space Research Institute,
the Funda\c{c}\~{a}o para a Ci\^{e}nca e a Tecnologia, the Ministerio
de Economia y Competitividad, the National Aeronautics and Space
Administration, the Netherlandse Onderzoekschool Voor Astronomie, the
Norvegian Space Center, the Romanian Space Agency, the United Kingdom
Space Agency and the University of Helsinki.  \bibliographystyle{mn2e}
\bibliography{euclid}

\appendix\section{The \emph{Euclid} spectroscopic survey}\label{app:spec}

\begin{figure}
\begin{center}
\begin{minipage}{0.5\textwidth}
\resizebox{\hsize}{!}{\includegraphics{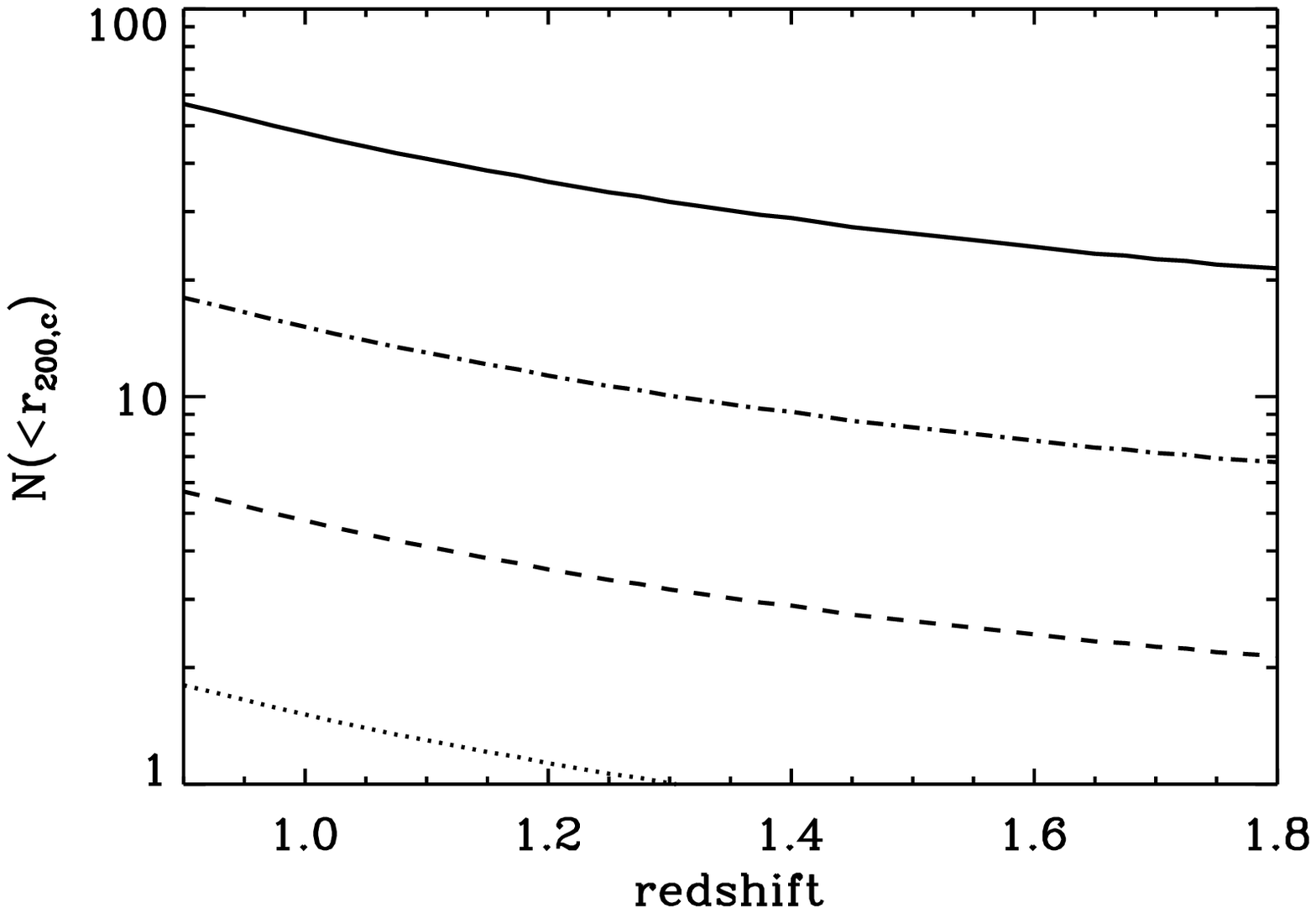}}
\end{minipage}
\end{center}
\caption{Number of cluster galaxies with spectroscopic
redshifts within $r_{200,c}$ expected in the \eu\ survey, as a
function of redshift for clusters of different masses, $\log
(M_{200,c}/M_{\odot})=15.0, 14.5, 14.0, 13.5$, solid, dot-dashed,
dashed, dotted lines, respectively. These numbers are for the case of
an evolving H$\alpha$ luminosity function also beyond $z=1.3$,
i.e. they correspond to the solid blue curve in
Fig.~\ref{f:selfuncspec} (top panel).}
\label{f:nzr200}
\end{figure}

\begin{figure}
\begin{center}
\begin{minipage}{0.5\textwidth}
\resizebox{\hsize}{!}{\includegraphics{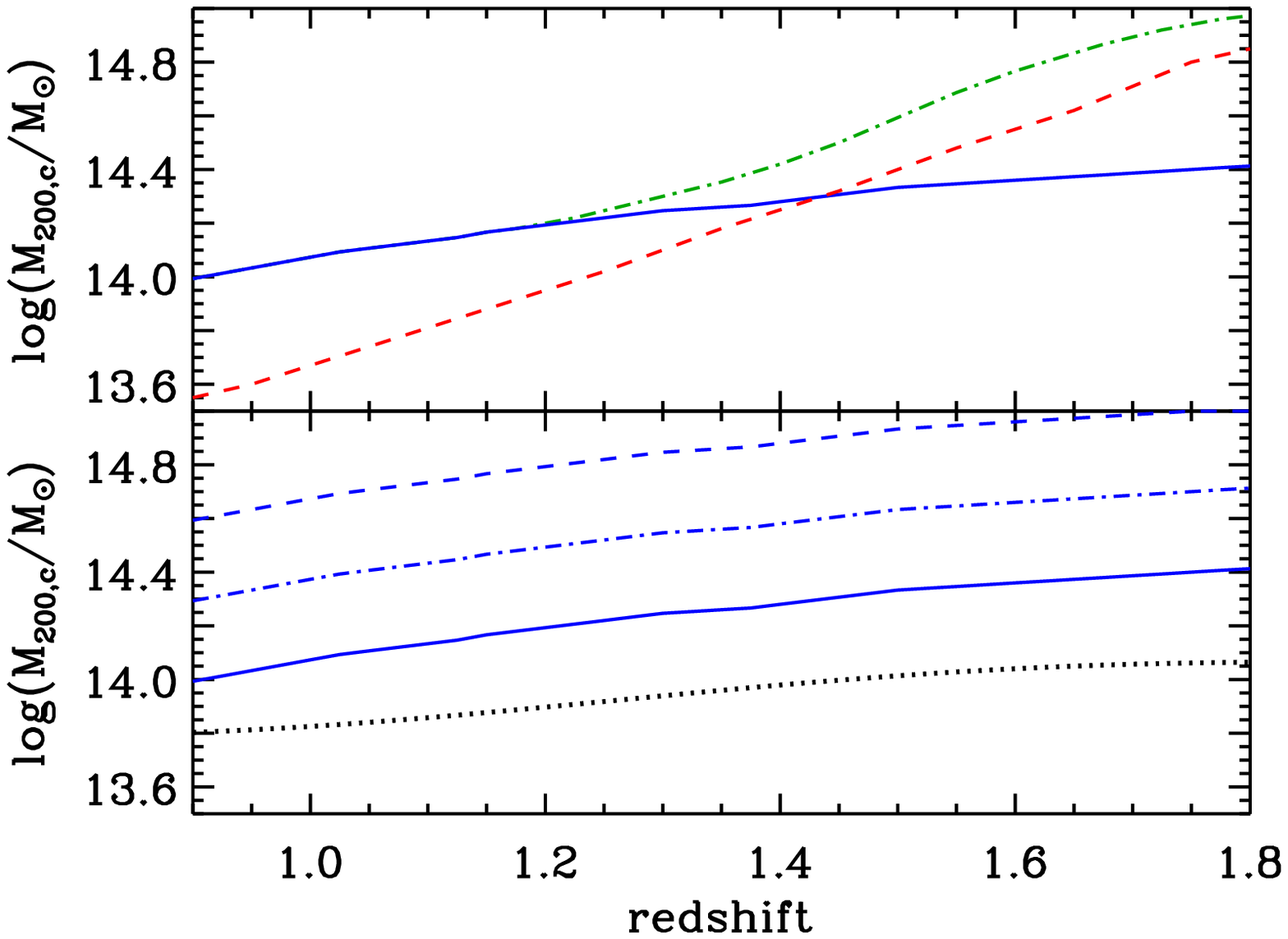}}
\end{minipage}
\end{center}
\caption{Selection function for the \eu\ spectroscopic survey.
In the top panel the solid blue curve indicates the selection function
for clusters with $5$ galaxies with measured spectroscopic redshift
within $\rtwo$. This curve depends on the assumption that $L^{\star}$
continues to evolve beyond $z=1.3$ following the same evolution law
determined by \citet{Geach+10} for lower $z$.  The dash-dotted curve
depends instead on the assumption that there is no further evolution
of $L^{\star}$ beyond $z=1.3$, consistently with what is observed for
the field $\ha$ LF \citep{Geach+10}.  The dashed red curve is an
independent estimate based on the the number densities of H$\alpha$
field galaxies per redshift bin, estimated by Pozzetti et al. (in
prep.).  In the lower panel the solid, dash-dotted and dashed lines
show results for clusters with at least $5$, $10$, and $20$ galaxies, respectively,
with measured spectroscopic redshift within $\rtwo$, based on the
assumption that $L^{\star}$ continues to evolve beyond $z=1.3$. The
dotted line is the selection function for the \eu\ photometric survey
(from Fig.~\ref{f:selfuncphot}), shown as a reference.}
\label{f:selfuncspec}
\end{figure}

We use a procedure similar to the one described in Section
\ref{s:clsel} to determine the number of spectroscopic cluster
galaxies within $\rtwo$, as a function of both cluster mass and
redshift.  Since the \eu\ spectroscopic survey is flux-limited in the
$\ha$ line, we consider the cluster $\ha$ LF. There are not many
determinations of the cluster $\ha$ LF in the literature. We use the
results of
\citet[][for two nearby clusters, $z=0.02$]{IglesiasParamo+02}, \citet[][for a
$z=0.18$ rich cluster]{Balogh+02}, \citet[][for a $z=0.25$
cluster]{Umeda+04}, and \citet[][for a $z=0.4$ cluster]{Kodama+04}.

The redshift evolution of the cluster $\ha$ LF is (at best) poorly
constrained, hence we have to make several assumptions for its three
parameters, the characteristic luminosity $L^{\star}$, the
normalisation $\phi^{\star}$, and the faint-end slope $\alpha$.  We
consider two possible evolutions. In the first, we assume the
$z$-evolution of $L^{\star}$ to be the same as the one measured for
the field $\ha$ LF, i.e.  $L^{\star} \propto (1+z)^{3.1}$ for $z \leq
1.3$, and no further evolution at higher redshift \citep{Geach+10}. In
the second, we allow $L^{\star}$ to evolve at $z>1.3$ with the same
$z$-dependence established at lower redshifts. The second
scenario is based on the idea that the preferred sites for
galaxy star-formation 
tends to shift to higher-density regions at higher
redshifts \citep{Elbaz+07}, even if the redshift at which this shift
occurs is not well constrained \citep{Ziparo+14}

The different cluster LFs we consider
have been determined for different overdensities, $\Delta$. To
evaluate the $\Delta=200$ value of $L^{\star}$ at $z=0$, we perform a
regression analysis between
$\mathrm{log}\,[L^{\star}/(1+z_\mathrm{c})^{3.1}]$ and
$\mathrm{log}\,\Delta$. We find $L^{\star}_{z=0}=3.8 \times 10^{41}$
erg~s$^{-1}$.  Similarly to what we did in Section \ref{s:clsel} for
the $\ks$ LF, we assume $\phi^{\star} \propto H^2(z)$.  We then take
the average of the $\phi^{\star}$ values obtained for the different
clusters, after rescaling them for the factor $200 \, H_0/[\Delta
H(z)]$, and find $\phi^{\star}_{z=0}=1.1$ Mpc$^{-3}$.  As for $\alpha$,
we fix it to the value $-0.7$ obtained for the two nearby clusters
by \citet{IglesiasParamo+02}, since the other clusters observations
were not deep enough to constrain the $\ha$ LF faint-end.

We convert the $\ha$ luminosities into fluxes using $f_{\ha} =
L_{\ha} / (2\times 4 \pi D_\mathrm{L,c}^2)$, where $D_\mathrm{L,c}$ is
the cluster luminosity distance and the factor $1/2$ accounts for the
average dust extinction \citep{Kodama+04}. By integrating the LF down
to the flux limit of the \eu\ spectroscopic survey ($3 \times
10^{-16}$ erg~s$^{-1}$~cm$^{-2}$), we finally obtain the expected
number density of galaxies within $\rtwo$.  By multiplying the number
density of galaxies within $\rtwo$ by the volume of the sphere of
radius $\rtwo$, we obtain the number of galaxies in a cluster with
$\ha$ flux above the \eu\ survey limit. Finally, we multiply
this number by the expected completeness of the spectroscopic survey, 
$\sim 80$ per cent.

In Fig. \ref{f:nzr200} we show the resulting estimates of the number
of cluster galaxies with spectroscopic redshifts within $r_{200,c}$, as
a function of redshift for clusters of different masses, for the case
of an evolving H$\alpha$ LF beyond $z=1.3$.  Note that only the
redshift range $0.9-1.8$ is shown, since this is the detectability
range of the H$\alpha$-line in the \eu\ survey, according to the
current design baseline\footnote{See the "\eu\ GC Interim Science
Review" by Guzzo \& Percival, at
http://internal.euclid-ec.org/?page\_id=714. Access restricted to
the \eu\ Consortium members.}.

We also consider the following, independent estimate of the cluster
selection function in the \eu\ spectroscopic survey. 
We use Pozzetti et al.'s (in prep.) estimates of the number densities
of H$\alpha$-emitting field galaxies per square degree and redshift
bin, that we convert to volume densities, $n_{fd}$. To estimate the
expected number density of H$\alpha$-emitting galaxies in a cluster, we
used $n_{cl}=n_{fd} b(z) \Delta \rho_c/\rho_m$, where
$\rho_c$ is the critical density and $\rho_m$ the mass density of the
Universe at any given redshift, $\Delta$ is the overdensity we want to
sample in the cluster, and $b(z)$ is the redshift-dependent bias
parameter that accounts for the different distribution of H$\alpha$
galaxies and the underlying matter distribution.  Taking $\Delta=200$,
the number of H$\alpha$ galaxies in a cluster of mass $M_{200,c}$ is
$N(\leq \rtwo)=(4 \pi / 3) r_{200,c}^3  n_{cl}$. We estimate
the bias $b(z)$ from the comparison of the real-space correlation
functions of matter and H$\alpha$ galaxies,
$b=(r_{0g}/r_{0m})^{-\gamma/2}$, where $\gamma$ is the slope of the
correlation function.  We use the correlation lengths of the diffuse
matter in our adopted cosmology, and those of H$\alpha$ galaxies with
luminosities corresponding to the \eu\ flux limit at any given
redshift \citep[taken from][]{Sobral+10}. We estimate $b(z=0.9)=1.9$
and $b(z=2.0)=3.5$, and interpolate $b(z)$ between these two values at
any redshift in the range 0.9--2.0.

In Fig. \ref{f:selfuncspec}, we show the limiting mass $\mtwo$ of a
cluster with at least $N_z$ galaxies with measured spectroscopic
redshift within $\rtwo$ as a function of the cluster redshift. This is
the selection function of clusters in the \eu\ spectroscopic survey,
in the sense that $N_z$ concordant redshifts within a region of
typical cluster size (i.e., $\rtwo$) are required to identify a
cluster. The three different estimates of the spectroscopic selection
function for clusters in the \eu\ survey are rather different, and
this reflects the current systematic uncertainties.  
From Fig. \ref{f:selfuncspec} (bottom panel), one can see that the
spectroscopic survey selection function is above the photometric
survey selection function. Hence, it will prove less efficient to
search for clusters in the \eu\ spectroscopic survey than in the
photometric survey. Data from the spectroscopic survey will still
be useful to confirm clusters detected in the photometric survey,
thus improving the reliability of the sample.

\section{Cluster mass calibration}\label{app:mass}

\begin{figure}
\includegraphics[width=0.47\textwidth]{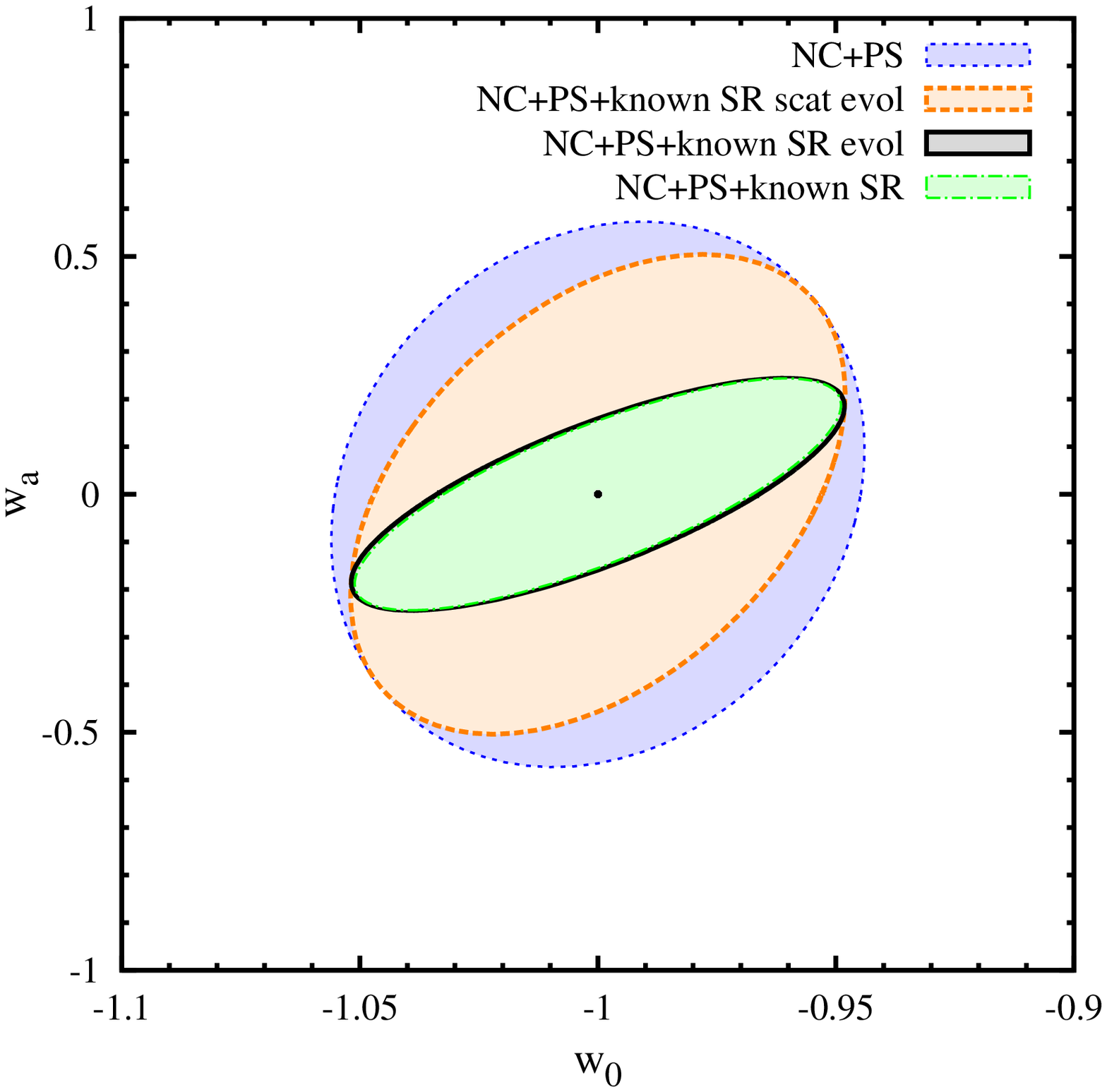}
\caption{Constraints at the $68$ per cent c.l. in the
  $w_a-w_0$ parameter plane. We show forecasts for the
  $N_{500,c}/\sigma_\mathrm{field}\ge 3$ \eu\ photometric cluster survey
  obtained by (i) combining the FM information for number counts and
  power spectrum (NC+PS; blue dotted contour), (ii) same as (i) but
  assuming perfect knowledge of the evolution of the scatter (see
  equation~\ref{eq:nuis}; orange dashed contour); (iii) same as (i) but
  assuming perfect knowledge of the evolution of both the scatter and
  the bias (black solid contour); (iv) same as (i) but assuming perfect
  knowledge of all the four nuisance parameters (green dash-dotted
  contour). The blue and green curves are the same of Fig.
\ref{fig:constrde},right panel. Note that the solid black and the green dashed
ellipses are almost coincident.}
\label{fig:nuis}
\end{figure}

The impact of nuisance parameters on cosmological constraints
from \eu\ photometric clusters is going to be quite significant. This
is especially true for the parameters directly related to the growth
of structure history like the matter power spectrum normalisation
$\sigma_8$, and for the CLP DE parameter $w_a$, that is particularly
sensitive to the level of knowledge of the scaling relation evolution.
In Fig.~\ref{fig:nuis}, we show how the cosmological constraints on
the DE equation of state depend on our knowledge of the scaling
relation. In particular, we show that strong constraints on the
evolution of the scatter and the mass bias, allow to greatly improve
the constraints on the DE EoS parameters. On the other hand, precise
knowledge of these parameters at $z=0$ is not of crucial importance,
as shown by the overlapping constraints in the $w_0,w_a$ plane
in the figure (solid black and dashed green ellipses).

To maximise the scientific return of the \eu\ galaxy cluster catalogue,
it is therefore very important to know the mass scaling relation in an as
much as possible precise and unbiased way. There are two avenues to
obtain this goal. The first one is to cross-correlate the \eu\ cluster
sample with samples obtained at different wavelengths by different
surveys. For instance, by the time
\eu\ will fly, the \emph{eRosita} full-sky X-ray cluster catalogue will
be available, and will provide an important contribution to the
cluster true mass estimation.  Other useful cluster catalogues will
include the SZ samples provided by the South-Pole Telescope (SPT), the
Atacama Cosmology Telescope (ACT), and \emph{Planck}.

\begin{figure}
\begin{center}
\begin{minipage}{0.5\textwidth}
\resizebox{\hsize}{!}{\includegraphics{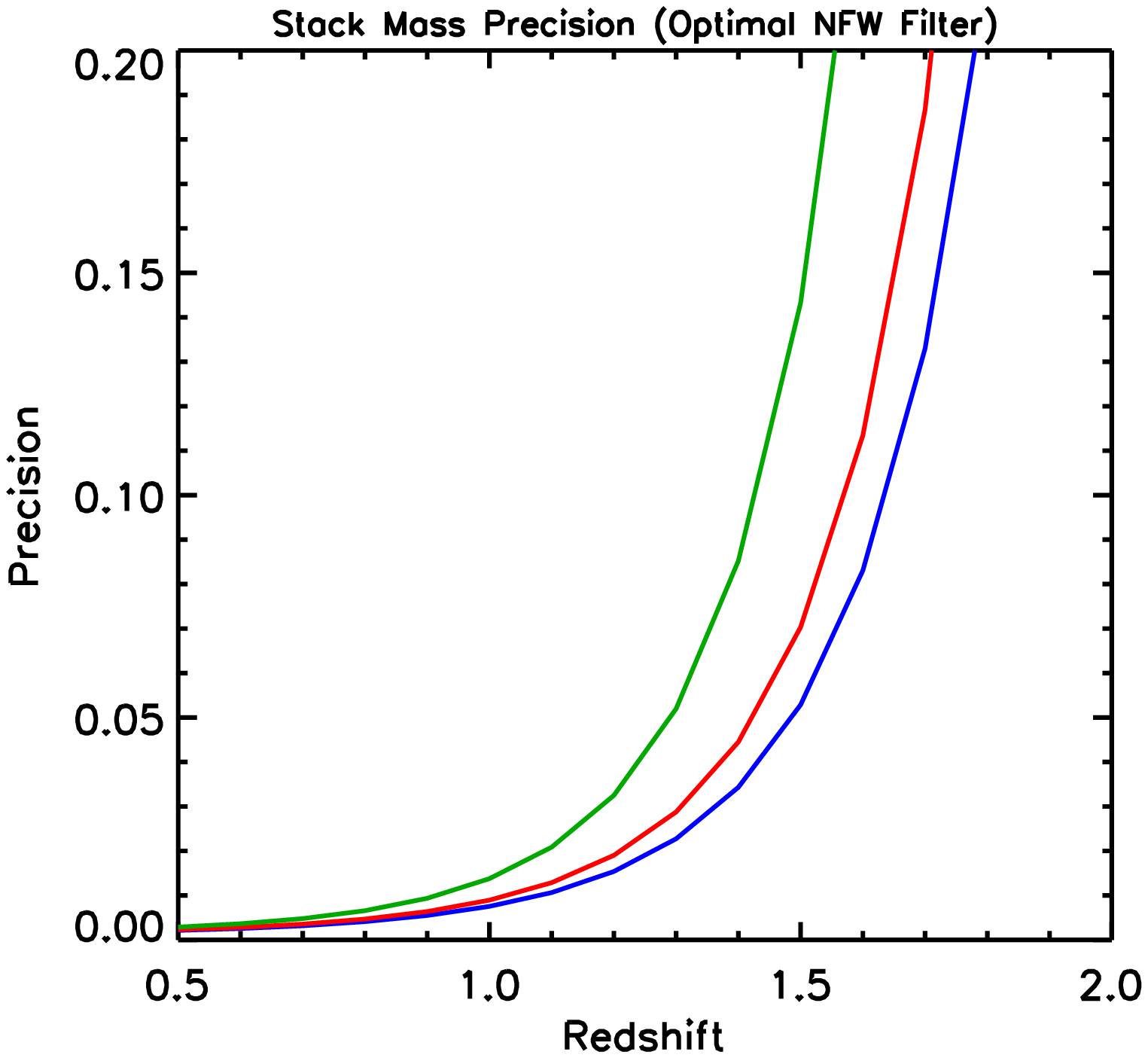}}
\end{minipage}
\end{center}
\caption{Calibrating cluster masses with gravitational shear.
The curves show the expected precision on the mean mass of clusters in
bins of $\Delta\log M_{200,c}=0.2$ and $\Delta z=0.1$, centred on masses
(from top to bottom) of $M_{200,c}=3\times 10^{14}\,\msun$ (green
curve), $2\times 10^{14}\,\msun$ (red), and $1.5\times 10^{14}\,\msun$
(blue).  We assume a lensing survey of 15,000 sq. deg.$^2$, the Tinker
mass function in the base $\Lambda$CDM \Planck-cosmology, and shape
noise with $\sigma=0.3$}
\label{f:stackprec}
\end{figure}

The second avenue, that represents the strength of the \eu\ mission,
consists in exploiting internal \eu\ data. Many photometrically
selected clusters will appear as signal-to-noise peaks in the
\eu\ full-sky cosmic shear maps. This weak gravitational lensing signal will
permit us to estimate the cluster masses without relying on
assumptions about dynamical equilibrium.  Although only the more
massive systems will permit individual mass measurements, we can
nevertheless statistically calibrate the normalisation of the cluster
scaling relations down to the lowest masses in the catalogue by
stacking. An example is given in Fig.~\ref{f:stackprec}, showing the
level of precision expected on the mean mass of stacked clusters.

We first measure the mass of individual clusters with a matched
filter, assuming that the mass density profile of all clusters follows an NFW
profile.  We then calculate the uncertainty on the mean mass of the
individual measurements in bins of mass ($\Delta \log M_{200,c}=0.2$)
and redshift ($\Delta z=0.1$).  This result depends on the number of
clusters expected in each bin, and for this purpose we have adopted
the \Planck\ cosmology \citep{planck13_16} and a \eu\ survey of 15,000
square degrees.  The figure only accounts for shape-noise, with
$\sigma=0.3$.

The three curves trace the precision on the mean for mass bins centred
at $M_{200,c}=3\times 10^{14}\,\msun$, $2\times 10^{14}\,\msun$, and
$1.5\times 10^{14}\,\msun$ (from top to bottom) as a function of
redshift.  We do better on the lower mass systems because their larger
number compensates for their lower individual signal-to-noise
measurements.  The figure demonstrates that \eu\ has the potential to
calibrate the mean mass, and hence scaling relations, to 1\%  out to
redshift unity, and to  10\% out to $z \lesssim 1.6$
for clusters of $M_{200,c}=1.5\times 10^{14}\,\msun$.
\newline

At the same time, the spectroscopic part of the
\eu\ survey will provide velocities for a few cluster members in 
each cluster detected with photometric data. Stacking these velocities
for many clusters in bins of richness and redshift will allow a
precise calibration of the velocity dispersion vs. richness relation,
and from this of the mass-richness relation.

\begin{figure}
\begin{center}
\begin{minipage}{0.5\textwidth}
\resizebox{\hsize}{!}{\includegraphics{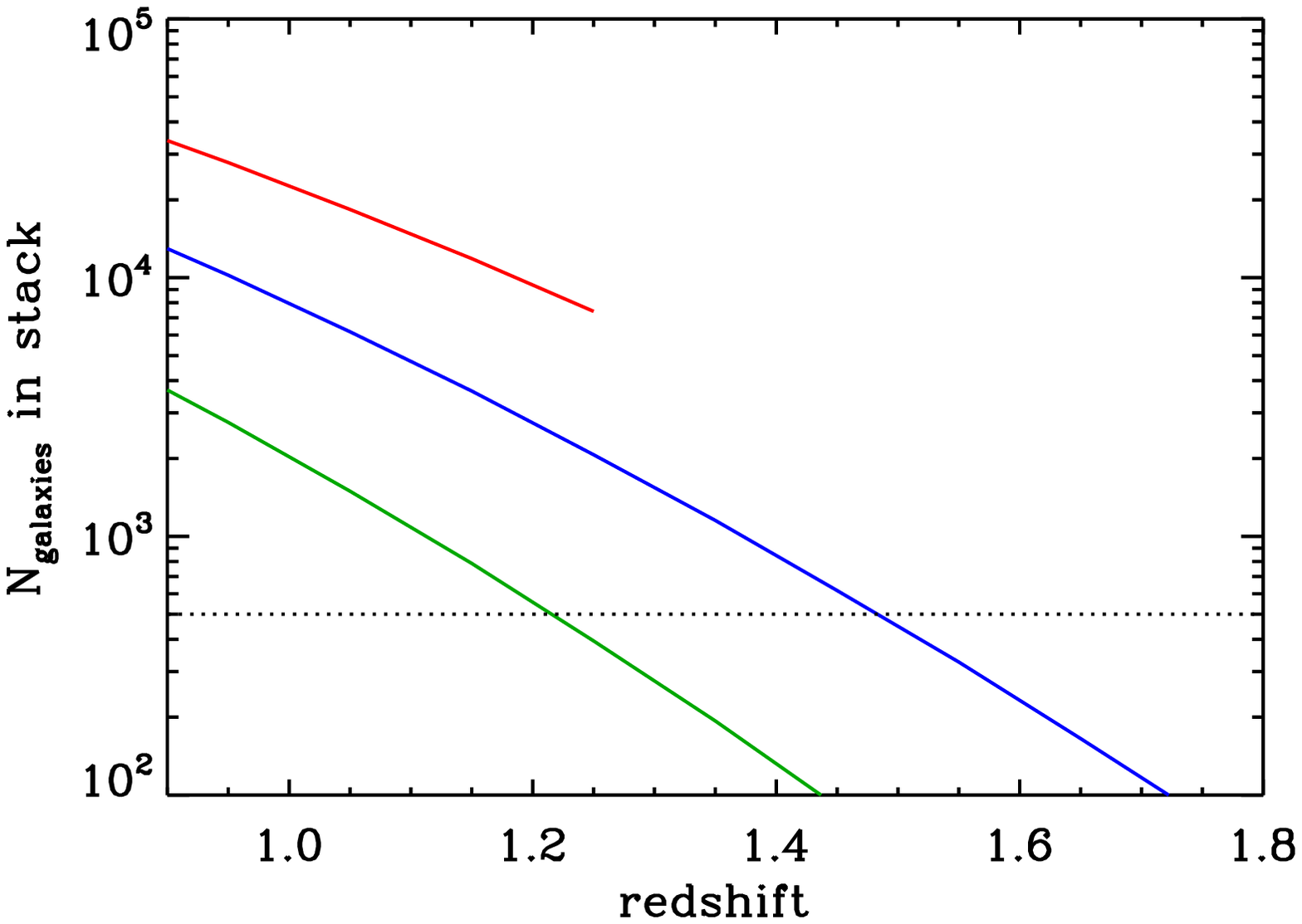}}
\end{minipage}
\end{center}
\caption{Calibrating cluster masses with spectroscopy.
The curves show the number of cluster galaxies with redshifts available
in stacks of clusters in bins of $\Delta \log M = 0.2$ and $\Delta z =
0.1$, as a function of redshift, for central values of the mass bins
of $\log M_{200,c}/M_{\odot}=14.2, 14.4, 14.6$ (red, blue, green curves,
respectively). The estimate is done only for clusters with a mass limit above
that required for a minimum of 5 members with redshift -- see
Fig. \ref{f:selfuncspec} bottom panel. This requirement restricts
the curve for $\log M/M_{\odot}=14.2$ to $z \leq 1.25$. The dotted
line shows the value of 500 galaxies as a reference.}
\label{f:stackz}
\end{figure}

In Fig.~\ref{f:stackz}, we show the number of spectroscopic cluster
members that will be available for stacks of clusters of given mass in
bins of $\Delta z = 0.1$ and $\Delta \log M_{200,c} = 0.2$ (even if, in
reality, the stacking procedure will be based on mass proxies, such as
richness).  These numbers are evaluated using the spectroscopic
selection function (bottom panel of Fig.~\ref{f:selfuncspec}), and
the expected number of clusters above a given mass in our adopted
cosmology, by considering only clusters with at least 5 members with
redshifts.  In the figure we show the predictions for three cluster
masses, $\log M_{200,c}/M_{\odot}=14.2, 14.4, 14.6$. The curve for $\log
M_{200,c}/M_{\odot}=14.2$ is limited to $z \leq 1.25$ because of our choice of
considering only clusters with $N_z \geq 5$.  Note that
the curve for $\log M_{200,c}/M_{\odot}=14.0$ (not shown) would be limited to
$z \leq 1$ (and it would be lie in between those for 14.2 and
14.4). 

From the analysis of \citet{BI06.1} we find that the statistical noise
in the velocity dispersion estimate of a sample of $\sim 500$ cluster
members is $\sim 9$ per cent, which translates into a $\sim 27$ per
cent statistical noise in the mass estimate. A similar figure has been
obtained by \citet{MBB13} when using the full velocity distribution to
constrain cluster masses. The value of 500 is displayed in
Fig.~\ref{f:stackz}, and it shows that a very precise spectroscopic
calibration of cluster masses will be possible for stacks of clusters
with $14.2 \leq \log M_{200,c}/M_{\odot} \leq 14.6$ over the redshift
range $0.9 \leq z \leq 1.2$, and even beyond that ($z \lesssim 1.5$)
for clusters with masses $\log M_{200,c}/M_{\odot} \simeq
14.4$. Spectroscopic calibration of cluster masses at higher redshifts
will be feasible with reduced precision, but lack of statistics will
hamper cluster mass calibration at $\log M_{200,c}/M_{\odot} <
14.2$. \newline

The wide \eu\ survey will allow precise calibration of the
mass-observable relation out to $z \lesssim 1.6$, using gravitational
lensing and spectroscopy. The deep \eu\ survey will allow to extend
this calibration to even higher redshifts, although with a much more
limited statistics on the number of clusters.  Overall, by
combining \eu\ internal mass calibration with the cross correlation
with external SZ and X-ray surveys, we should be able to significantly
mitigate the degrading effect of the nuisance parameters on
cosmological constraints.

\end{document}